\documentclass[letterpaper, 10 pt, conference]{IEEEtran}
\usepackage[latin1]{inputenc}
\usepackage[numbers, square, comma, compress]{natbib}
\usepackage[english]{babel}
\usepackage{dsfont}
\usepackage{setspace}
\usepackage{amsfonts}
\usepackage{amssymb}
\usepackage{amsmath}
\usepackage{mathrsfs}
\usepackage{xcolor}
\usepackage{graphicx}
\usepackage{mathdots}
\usepackage{float}
\usepackage{dblfloatfix}
\usepackage{stmaryrd}
\usepackage{epstopdf}
\usepackage[colorinlistoftodos]{todonotes}
\usepackage[ruled,vlined]{algorithm2e}
\usepackage{verbatim}
\usepackage{hyperref}
\usepackage{color}
\usepackage{lipsum}
\usepackage{mathtools}
\usepackage{cuted}
\usepackage{paralist}

\usepackage{geometry}
\usepackage{tikz}
\usetikzlibrary{shapes,arrows}
\usetikzlibrary{calc}

\pdfoutput=1

\newtheorem{teo}{Theorem}
\newtheorem{prop}{Proposition}
\newtheorem{lem}{Lemma}

\newtheorem{oss}{Remark}
\newtheorem{defi}{Definition}

\newcounter{mytempeqncnt}

\renewcommand{\IEEEQED}{\IEEEQEDopen}

\makeatletter
\renewcommand*\env@matrix[1][*\c@MaxMatrixCols c]{%
  \hskip -\arraycolsep
  \let\@ifnextchar\new@ifnextchar
  \array{#1}}
\makeatother

\IEEEoverridecommandlockouts

\title{
\raisebox{6.3mm}{\strut} 
Strong Coordination over Noisy Channels with Strictly Causal Encoding}

\author{\IEEEauthorblockN{Giulia Cervia\IEEEauthorrefmark{1}, Laura Luzzi\IEEEauthorrefmark{1}, Ma\"{e}l Le Treust\IEEEauthorrefmark{1} and Matthieu R. Bloch\IEEEauthorrefmark{3}
\thanks{Ma\"{e}l Le Treust gratefully acknowledges the supports of DIM-RFSI under grant EX032965, and of Labex MME-DII (ANR11-LBX-0023-01). The authors thank SRV ENSEA for financial support for the visit of M. R. Bloch in 2017.}
}
\IEEEauthorblockA{\IEEEauthorrefmark{1} ETIS UMR 8051, Universit\'{e} Paris Seine, Universit\'{e} Cergy-Pontoise, ENSEA, CNRS, Cergy, France. \\
email: \{giulia.cervia, laura.luzzi, mael.le-treust\}@ensea.fr}
\IEEEauthorblockA{\IEEEauthorrefmark{3}School of Electrical and Computer Engineering, Georgia Institute of Technology, Atlanta, Georgia\\
email: matthieu.bloch@ece.gatech.edu} 
}

\begin{document}
\bstctlcite{IEEEexample:BSTcontrol}
\IEEEoverridecommandlockouts
\maketitle

 \begin{abstract} 
 
We consider a network of two nodes separated by a noisy channel, 
in which the input and output signals have to be coordinated with the source and its reconstruction.
In the case of strictly causal encoding and non-causal decoding, we prove inner and outer bounds for the strong coordination region and show that the inner bound is achievable with polar codes.
 \end{abstract}
 
\section{Introduction}
While communication networks have traditionally been designed to reliably convey information, 
modern decentralized networks are introducing new challenges. 
More than communication by itself, what is crucial for the next generation of networks is to ensure the \emph{cooperation} and \emph{coordination} of the 
constituent devices, viewed as autonomous decision makers. 
The devices have to adapt their behavior to the state of the environment 
and to the actions of other devices, which may not be known by all players, creating information asymmetries; coordination is meant in the broad sense of enforcing a 
joint behavior of the devices through communication to resolve such asymmetries.

More specifically, we quantify coordination in terms of how well we can approximate a target joint distribution between the actions and signals of the devices. 
In particular, \emph{empirical coordination} requires the joint histogram of actions and signals to approach a target distribution, while \emph{strong coordination} 
requires their joint distribution to converge in total variation to an i.i.d. target distribution~\cite{cuff2010}.

In this work, we consider a two-node network with an information source and a noisy channel in which the input and output signals should be strongly coordinated with the source and the reconstruction. 
This scenario presents two conflicting goals: the encoder needs to convey a message to the decoder to coordinate the actions,
while simultaneously coordinating the signals coding the message. The two nodes are assisted in their task by a shared source of randomness. 
The case in which the encoder and the decoder are both non-causal has already
been considered in~\cite{Cervia2017,Cervia2018journal} but the problem of finding the coordination region is still open.
We focus here on the setting in which the encoder is strictly causal,
which has the benefit of shortening the transmission delay.
 
In~\cite{cuff2011hybrid} the authors provide a characterization of the empirical coordination region when the encoder is strictly causal.
In~\cite{Cervia2017gretsi}, we proposed an
explicit polar coding scheme that achieves this region. 
In this paper, we provide an inner and an outer bound for the strong coordination region and show that the inner bound is achievable with polar codes. 
Although the achievability techniques are similar to the ones used 
in~\cite{Cervia2018journal}, the strictly causal nature of the encoder requires a more subtle random coding scheme with a block-Markov structure. 

The remainder of the paper is organized as follows.
$\mbox{Section \ref{sec: prel}}$ introduces the notation, $\mbox{Section \ref{sec: model}}$ 
describes the model under investigation and 
states the main result.
$\mbox{Section \ref{inner}}$ proves an inner bound by proposing 
a random binning scheme and a random coding scheme that have the same statistics and $\mbox{Section \ref{outerpart1}}$ proves an outer bound.
The two bounds match, except for the bound on the minimal rate of common randomness, and closing the gap between the two regions remains an open problem.
Finally, we provide an explicit polar code construction achieving 
the inner bound in the appendix.

\section{Preliminaries}\label{sec: prel}

We define the integer interval $\llbracket a,b \rrbracket$ as the set of integers between $a$ and $b$.
Given a random vector $X^{n}:=$ $(X_1, \ldots, X_{n})$, we note $X^{i}$ the first $i$ components of $X^{n}$, 
$X_{\sim i}$ the vector $(X_j)_{j \neq i}$, $j\in \llbracket 1,n \rrbracket $, where the component $X_i$ has been removed and $X[A]$ the vector $(X_j)_{j \in A}$, 
$A \subseteq \llbracket 1,n \rrbracket $. Given two random vectors $A$ and $B$, $A \perp B$ indicates that $A$ and $B$ are independent.
We denote with $Q_A$ the uniform distribution over $\mathcal A$.
We note $\mathbb V (\cdot , \cdot)$ and $\mathbb D (\cdot \Arrowvert \cdot)$ the variational distance and the Kullback-Leibler divergence between two distributions.
The notation $f(\varepsilon)$ denotes a function which tends to zero as $\varepsilon$ does, 
and the notation $\delta(n)$ denotes a function which tends to zero exponentially as $n$ goes to infinity.

We now state some useful results.

\begin{lem}[$\mbox{\cite[Lemma 17]{cuff2009thesis}}$]\label{cuff17}
\begin{small}
 $\! \!\!\!\mathbb V (P_A, \! \hat P_A)\!\!= \!\!\mathbb V (P_A \! P_{B|A},\! \hat P_A \! P_{B|A}).$
\end{small}
\end{lem}

\vspace{0,2cm}
\begin{lem}\label{lemkl}
\begin{small}
 $\mathbb D \left(P_A \Arrowvert \hat P_A \right)= \mathbb D \left(P_AP_{B|A} \Arrowvert \hat P_AP_{B|A}\right).$
\end{small}
\end{lem}

\begin{center}
\begin{figure}[t]
 \centering
 \includegraphics[scale=0.21]{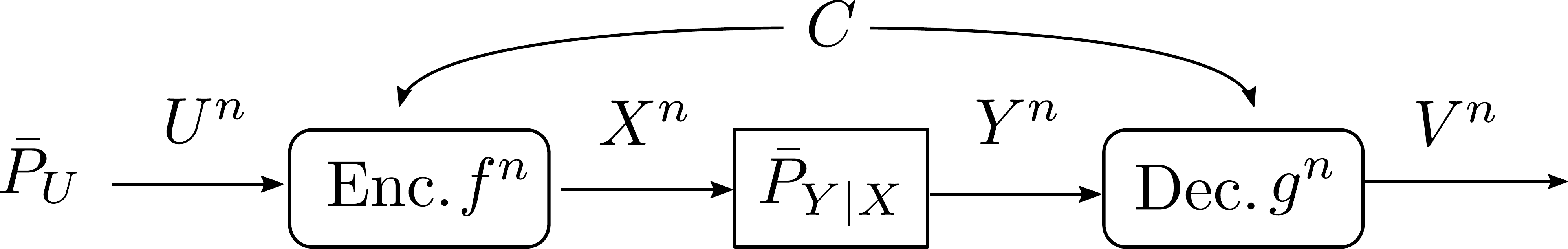}
\caption{Coordination of signals and actions for a two-node network with a noisy channel with strictly causal encoder and non-causal decoder.}
\label{fig: coordisit}
\end{figure}
\end{center}

\begin{lem}[$\mbox{\cite[Lemma 2.7]{csiszar2011information}}$]\label{csi2.7}
Let $P$ and $P'$  two probability mass functions on $\mathcal A $  
such that 
$\mathbb V(P,P') = \varepsilon\leq 1/2$, then
\begin{equation*}
 \lvert H(P)-H(P') \rvert \leq  \varepsilon \log{\frac{\lvert \mathcal A \rvert}{\varepsilon}}.
\end{equation*}
\end{lem}

\begin{lem}[$\mbox{\cite[Lemma 4]{yassaee2014achievability}}$]\label{lem4}
If \begin{small}
    $ \mathbb V (P_{A^{n}} P_{B^{n}|A^{n}} ,\!P'_{A^{n}} P'_{B^{n} |A^{n}} ) \!\!= \varepsilon$, 
   \end{small}then there exists $\mathbf a \in \mathcal A^{n}$ such that
 \begin{equation*}
 \mathbb V \left(P_{B^{n}|A^{n}= \mathbf a},P'_{B^{n} |A^{n}= \mathbf a} \right) = 2 \varepsilon.
 \end{equation*}
\end{lem}

\begin{lem}[$\mbox{\cite[Lemma 6]{Cervia2018journal}}$]\label{lemmit}
Let $P_{A^{n}}$ such that 
$\mathbb V\left(P_{A^{n}}, \bar P_{A}^{\otimes n}\right)$  is smaller than $ \varepsilon,$
then we have
\begin{equation*}
\sum_{t=1}^{n} I(A_t;A_{\sim t}) \leq n f(\varepsilon). 
\end{equation*}
\end{lem}

\section{System model and main result}\label{sec: model}

We  consider the model depicted in Figure \ref{fig: coordisit}. 
Two agents, the encoder and the decoder, wish to coordinate their behaviors, in the sense that the stochastic actions of the agents should follow a known and fixed joint distribution.
We suppose that the encoder and the decoder have access to a shared source of uniform randomness $C \in \llbracket 1,2^{nR_0} \rrbracket$.
Let $U^{n} \in \mathcal U^n $ be an i.i.d. source with distribution $\bar P_{U}$.
At time $i \in \llbracket 1,n\rrbracket$, 
the strictly causal encoder observes the sequence $U^{i-1} \in \mathcal U^n$,
common randomness $C$ and selects a signal $X_i= f_i(U^{i-1},  C)$, where 
$f_i: \mathcal U^{i-1}  \times  \llbracket 1,2^{nR_0} \rrbracket \rightarrow \mathcal X$
is a stochastic function.
The signal $X^{n}=(X_1, \ldots, X_n)$ is transmitted over a discrete memoryless channel $\bar P_{Y |X}$.
Upon  observing  $Y^{n}$ and the common randomness $C$, 
the decoder selects an action $V^{n} = g^n(Y^{n}, C)$, where
$g^n: \mathcal Y^n  \times \llbracket 1,2^{nR_0} \rrbracket \rightarrow \mathcal V^n$ is a stochastic  map. 
Let $f^n:=\{f_i\}_{i=1}^n$ for block length $n$. The pair $(f^n , g^n )$  constitutes a code. 
We introduce the definitions of achievability and of strong coordination in this setting.

\begin{defi}\label{definition sc ccord}
A pair $(\bar{P}_{UXYV}, R_0)$ is \emph{achievable} for \emph{strong coordination} if there exists a sequence $(f^n,g^n)$ of 
strictly causal encoders and non causal decoders with rate of common randomness $R_0$, such that for every $\varepsilon>0$ there exists
$n \in \mathbb N$ and a sufficiently long sub-sequence $(U^{\tilde n}, X^{\tilde n}, Y^{\tilde n},  V^{\tilde n})$
 with $\tilde n>(1-\varepsilon)n $ that satisfies
$$\lim_{n \to \infty} \mathbb V \left( P_{U^{\tilde n} X^{\tilde n} Y^{\tilde n}  V^{\tilde n}}, \bar{P}_{UXYV}^{\otimes \tilde n} \right)=0$$
where $P$ is the joint distribution induced by the code.
The \emph{strong coordination region} $\mathcal{R}$ is the closure of the set of achievable pairs 
$(\bar P_{UXYV}, R_0)$\footnote{To avoid boundary complications,  we define the achievable region as the closure of the set of achievable rates and distributions
as in \cite{cuff2010}. For a careful discussion on the boundaries the region, see \cite[Section VI.D]{cuff2013distributed}.}.
\end{defi}

\begin{oss}
The definition for strong coordination in this setting is slightly different from the definition of strong coordination with non-causal encoder and
decoder in~\cite{cuff2010,Cervia2018journal}, which for the strictly-causal encoder would be satisfied only by trivial distributions since the last block of the source will never be observed by the encoder.
Here, we avoid this issue by losing coordination in a negligible fraction of time slots.
\end{oss}

The problem of characterizing the strong
coordination region is still open, but we establish the following inner and outer bounds.
\begin{teo} \label{teostrictly}
Let $\bar P_{U}$ and $\bar P_{Y|X}$ be the given source and channel parameters, then  
$\mathcal R_{\text{in}} \subseteq \mathcal{R}\subseteq \mathcal R_{\text{out}}$ 
{\allowdisplaybreaks
\begin{align} 
\!&\mathcal R_{\text{in}} \!\!:=\! \! \begin{Bmatrix}[c|l]
& \! \bar{P}_{UXYV}=\bar P_{U} \bar P_{X} \bar P_{Y|X} \bar P_{V|UXY} \! \\
& \! \exists \mbox{ } W \mbox{ taking values in $\mathcal W$}\!\\ 
\!\!(\bar P_{UXYV}, R_0) \! \! &\! \bar{P}_{UXYVW}=\!\\
&\! \bar P_{U} \bar P_{X} \bar P_{W|UX} \bar P_{Y|X} \bar P_{V|WY}\!\\
& \!I(WX;U) \leq I(WX;Y) \!\\
& \!R_0 \geq I(W; UXV|Y)+H(X|WY)\!
\end{Bmatrix}\label{eq: innerbound}\\
& \mathcal R_{\text{out}} \!:=\!\! \begin{Bmatrix}[c|l]
& \bar{P}_{UXYV}=\bar P_{U} \bar P_{X} \bar P_{Y|X} \bar P_{V|UXY}  \\
&\exists \mbox{ } W \mbox{ taking values in $\mathcal W$}\\ 
& \bar{P}_{UXYVW}=\\
\!\!(\bar P_{UXYV}, R_0) \! \!& \bar P_{U} \bar P_{X} \bar P_{W|UX} \bar P_{Y|X} \bar P_{V|WY}\\
& I(WX;U) \leq I(WX;Y) \\
& R_0 \geq I(W; UXV|Y)\\
&\lvert \mathcal W \rvert \leq \lvert \mathcal U \times  \mathcal X \times \mathcal Y \times {\mathcal V} \rvert+4
\end{Bmatrix}.\label{eq: outerbound}
\end{align}} 
\end{teo}

\begin{teo}\label{polarregion}
The region $\mathcal R_{\text{in}}$ defined in \eqref{eq: innerbound} is achievable using polar codes, provided there exists an error-free channel
of negligible rate between the encoder and decoder.
\end{teo}

 \begin{oss}
By the chain rule, we have
\begin{itemize}
\item[\textbullet]$I(XW;U)= I(W;U|X)+I(X;U)=I(W;U|X)$ since $U$ and $X$ are independent;
\item[\textbullet] $I(XW;Y)= I(W;Y|X)+I(X;Y)=I(X;Y)$ because of the Markov chain $W-X-Y$.
\end{itemize}
Hence the condition $I(WX;U) \leq I(WX;Y)$ in \eqref{eq: innerbound} and \eqref{eq: outerbound} is equivalent to $I(W;U|X)\leq I(X;Y)$.
\end{oss}

\paragraph*{Comparison with empirical coordination}
For empirical coordination, \cite[Theorem 3]{cuff2011hybrid} gives the following characterization of the region with strictly causal encoding: 
 {\allowdisplaybreaks
\begin{equation} \label{eq: regioncuffschieler}
\!\mathcal R_{\text{emp}}\! :=\!\!  \begin{Bmatrix}[c|l]
& \!\bar{P}_{UXYV}=\bar P_{U} \bar P_{X} \bar P_{Y|X} \bar P_{V|UXY} \! \\
\! \bar P_{UXYV} \!\!& \! \exists \mbox{ } W \mbox{ taking values in $\mathcal W$}\!\\
&\! \bar{P}_{UXYVW}\!=\! \bar P_{U} \bar P_{X} \bar P_{W|UX} \bar P_{Y|X} \bar P_{V|WY}\!\\
&\! I(WX;U) \leq I(WX;Y) \!
\end{Bmatrix}\!\!.
\end{equation}}Observe that in  $\mathcal R_{\text{in}}$ and $\mathcal R_{\text{out}}$ the decomposition of the joint distribution and the information constraints 
are the same as in $\mathcal R_{\text{emp}}$, but for strong coordination a positive rate of common randomness is also necessary.
This is consistent with the conjecture, stated in \cite{cuff2010}, that with enough common randomness
the strong coordination capacity region is the same as the empirical coordination capacity region for any specific network setting.

\section{Achievability proof of Theorem \ref{teostrictly}}\label{inner}
The key idea of the achievability proof is to define a random binning for the target joint distribution, and a random coding scheme, each of
which induces a joint distribution, and to prove that the two schemes have almost the same statistics. 
The proof uses the same techniques as in \cite{haddadpour2017simulation} 
inspired by \cite{yassaee2014achievability},  to deal with the strictly causal encoder, a block Markov structure is required for the random coding scheme.  
Before defining the coding scheme, we state some results that we use to prove the inner bound.

The following lemma is a consequence of the Slepian-Wolf Theorem.  
\begin{lem}[Source coding with side information at the decoder $\mbox{\cite[Theorem 10.1]{elgamal2011nit}}$ ]\label{lem1}
 Given a discrete memoryless source $(A^{n},B^{n})$, where $B^{n}$ is side information available at the decoder,  
 let $\varphi_n: \mathcal A^n \to \llbracket 1,2^{nR} \rrbracket$  be a uniform random binning of $A^n$, 
  and let $C:= \varphi_n(A^{n})$.
Then if $R>H(A|B)$, the decoder can recover $A^{n}$ from $C$ and $B^{n}$ with:
\begin{equation*}
 \mathbb E_{\varphi_n} [\mathbb P \{\hat A^n \neq A^n \} ]\leq \delta(n).
\end{equation*}
\end{lem}

 \begin{lem}[Channel randomness extraction for discrete memoryless sources and channels]\label{1.4.2}
  Let $A^n$ with distribution $P_{A^n}$ be a discrete memoryless source and $P_{B^n|A^n}$ a discrete memoryless channel. 
  Let $\varphi_n: \mathcal B^n \to \llbracket 1,2^{nR} \rrbracket$  be a uniform random binning of $B^n$, 
  and let $K:= \varphi_n(B^{n})$.
Then if $R \leq H(B|A)$, there exists a constant $\alpha > 0$ such that
  \begin{equation}\label{eq1lem1.4.2}
  \mathbb E_{\varphi_n} [ \mathbb D (P_{A^nK} \Arrowvert P_{A^n} Q_K )] \leq 2^{-\alpha n}.
  \end{equation}
  \end{lem}

We omit the proof of Lemma \ref{1.4.2} as it follows directly from the discussion in \cite[Section III.A]{pierrot2013joint}.

\subsection{Random binning scheme}\label{rb2}

Assume that the sequences $U^n$, $X^n$, $W^n$, $Y^n$ and $V^n$ are jointly i.i.d. with distribution
\begin{equation}\label{iiddistr}
 \bar P_{U^n} \bar P_{X^n} \bar P_{W^n| U^n X^n} \bar P_{Y^n|X^n} \bar P_{V^n|W^n Y^n}.
\end{equation} 
First, we consider two uniform random binnings for $X^n$:
\begin{itemize}
\item $M_1 = \varphi_1(X^n)$, where  $\varphi_1: \mathcal{X}^n \to \llbracket 1,2^{nR_1}\rrbracket$,
\item $M_2 = \varphi_2(X^n)$,  $\varphi_2: \mathcal{X}^n \to \llbracket 1,2^{n R_2}\rrbracket$.
\end{itemize}The rates $R_1$ and $R_2$ are chosen as follows:
\begin{itemize}
 \item $R_1+R_2<H(X)$, so that by Lemma \ref{1.4.2} there exists one binning $(\varphi'_1, \varphi'_2)$ of $X$ such that $M_1$ and $M_2$ are almost uniform and almost independent of each other;
 \item $R_1>H(X|Y)$, so that by Lemma \ref{lem1} there exists one binning $ \varphi'_1$ of $X$ such that it is possible to reconstruct $X$ from $Y$ and $M_1$ with 
high probability using a Slepian-Wolf decoder via the conditional distribution $P^{\text{SW}}_{\hat X^n |M_1  Y^n}$;
\end{itemize}where we can use the same binning $\varphi'_1$ for both conditions, as proved in \cite[Remark 7]{Cervia2018journal}.

Then, we consider the following uniform random binnings for $W^n$:
\begin{itemize}
\item $M_3 = \varphi_3(W^n)$, $\varphi_3: \mathcal{W}^n \to \llbracket 1,2^{n R_3}\rrbracket$,
\item $M_4 = \varphi_4(W^n)$,  $\varphi_4: \mathcal{W}^n \to \llbracket 1,2^{n R_4}\rrbracket$,
\item $F = \psi(W^n)$,  $\psi: \mathcal{W}^n \to \llbracket 1,2^{n \tilde R}\rrbracket$,
\end{itemize}where the rates $R_3$, $R_4$ and $\tilde R$ are chosen as follows:
\begin{itemize}
 \item $R_3+ \tilde R<H(W|XU)$, so that by Lemma \ref{1.4.2} there exists one binning $(\varphi'_3, \psi')$ of $W$ such that $M_3$ and $F$ are almost uniform and almost independent of $X$ and $U$;
 \item $R_3+R_4 +\tilde R>H(W|X)$, so that by Lemma \ref{lem1} there exists one binning $(\varphi'_3,\varphi'_4, \psi')$ of $W$ such that it is possible to reconstruct $W$ from $X$ and $(M_3, M_4,F)$ with 
high probability using a Slepian-Wolf decoder via the conditional distribution 
$P^{\text{SW}}_{\hat W^n |M_3 M_4 F X^n}$;
\end{itemize}and we can use the same binning $(\varphi'_3, \psi')$ for both conditions, as proved in \cite[Remark 7]{Cervia2018journal}.
This defines a joint distribution:
\begin{align*}
P^{\text{RB}}:=
  &  \bar P_{U^n} \bar P_{X^n} \bar P_{W^n|U^n X^n}  \bar P_{M_1|X^n} \bar P_{M_2|X^n}   \bar P_{M_3|W^n} \stepcounter{equation}\tag{\theequation}\label{rbi}\\
  &\bar P_{M_4|W^n} \bar P_{F|W^n} \bar P_{Y^n|X^n} \bar P_{V^n|W^n Y^n}.
\end{align*}
In particular, the conditional distributions $P^{\text{RB}}_{M_4|M_3 X^n U^n}$, $P^{\text{RB}}_{W^n|M_3 M_4 F X^n}$ 
and $P^{\text{RB}}_{X^n|M_1 M_2 M_3 F}$ are well-defined.

\subsection{Random coding scheme}\label{rc2}

In this section we follow the approach in \cite[Section IV.E]{yassaee2014achievability} and \cite{haddadpour2017simulation}.
Suppose that encoder and decoder have access to extra randomness $F$, where  $F$ is generated uniformly at
random in $\llbracket 1,2^{n \tilde R} \rrbracket$ with distribution $Q_F$ independently of the rest of the common randomness.

\subsubsection{Encoder}
We use a chaining construction over $k$ blocks of length $n$ in which
the encoder observes  $U^{n}_{(1:k)}:=(U^{n}_{(1)}, \ldots, U^{n}_{(k)})$, where  $U^{n}_{(i)}$
for $i \in \llbracket 1,k \rrbracket$ are $k$ blocks of the source.
The encoder has access to common randomness $(M_{1,(1:k)},$ $ M_{3,(1:k)}, F_{(1:k)}, K_{(2:k)})$ and the block-Markov scheme proceeds as follows. 

\begin{figure*}[!t]
\normalsize
\setcounter{mytempeqncnt}{\value{equation}}
\setcounter{equation}{7}
\begin{align*}
P^{\text{RC}}_{(i)}: =&P^{\text{RC}}_{{(U^n  X^n \hat X^n Y^n V^n W^n  M_{1} M_{2} M_{3} M_{4} \hat M_4 F)}_{(i)}} \\
=& \bar P_{U^n} (\mathbf u_{(i)})  Q_{M_1} ( \mathbf m_{1,(i)}) Q_{M_2} ( \mathbf m_{2,(i)}) Q_{M_3} ( \mathbf m_{3,(i)}) Q_{F} ( \mathbf f_{(i)}) 
P^{\text{RB}}_{X^n|M_1 M_2 M_3 F} (\mathbf x_{(i)}| \mathbf m_{1,(i)}, \mathbf m_{2,(i)}, \mathbf m_{3,(i)},\mathbf f_{(i)} )\\
&   P^{\text{RB}}_{M_4|M_3 X^n U^n}(\mathbf m_{4,(i)}| \mathbf m_{3,(i)}, {\mathbf x}_{(i)}, \mathbf u_{(i)}) \bar P_{Y^n|X^n}(\mathbf y_{(i)}| \mathbf x_{(i)} ) P^{\text{SW}}_{\hat X^n |M_1  Y^n} (\hat{\mathbf x}_{(i)}|  \mathbf m_{1,(i)}, \mathbf y_{(i)} ) P^{\text{RC}}_{\hat M_4}(\hat{\mathbf m}_{4,(i)}) \stepcounter{equation}\tag{\theequation}\label{rci}\\
&  P^{\text{SW}}_{W^n |M_3 \hat M_4 F \hat X^n}  ({\mathbf w}_{(i)}|  \mathbf m_{3,(i)}, \hat{\mathbf m}_{4,(i)}, \mathbf f_{(i)}, \hat{\mathbf x}_{(i)} ) P^{\text{RC}}_{V^n| W^n  Y^n}  ({\mathbf v}_{(i)}| {\mathbf w}_{(i)}, \mathbf y_{(i)} ).\\
\setcounter{equation}{9}\\
\hat P^{\text{RC}}_{(i)}:=&\hat P^{\text{RC}}_{{(U^n  X^n Y^n V^n W^n  M_{1} M_{2} M_{3} M_{4}  F)}_{(i)}} \\
=&\bar P_{U^n} (\mathbf u_{(i)})  Q_{M_1} ( \mathbf m_{1,(i)}) Q_{M_2} ( \mathbf m_{2,(i)}) Q_{M_3} ( \mathbf m_{3,(i)}) Q_{F} ( \mathbf f_{(i)}) P^{\text{RB}}_{X^n|M_1 M_2 M_3 F} (\mathbf x_{(i)}| \mathbf m_{1,(i)}, \mathbf m_{2,(i)}, \mathbf m_{3,(i)},\mathbf f_{(i)} )\\
& P^{\text{RB}}_{M_4|M_3 X^n U^n}(\mathbf m_{4,(i)}| \mathbf m_{3,(i)},  {\mathbf x}_{(i)}, \mathbf u_{(i)})  \bar P_{Y^n|X^n}(\mathbf y_{(i)}| \mathbf x_{(i)} )\stepcounter{equation}\tag{\theequation}\label{rci tilde} \\
& P^{\text{SW}}_{W^n |M_3 M_4 F  X^n}  ({\mathbf w}_{(i)}|  \mathbf m_{3,(i)}, {\mathbf m}_{4,(i)}, \mathbf f_{(i)}, {\mathbf x}_{(i)} ) P^{\text{RC}}_{V^n| W^n  Y^n}  
({\mathbf v}_{(i)}| {\mathbf w}_{(i)}, \mathbf y_{(i)} ).
\end{align*}
\setcounter{equation}{\value{mytempeqncnt}}
\hrulefill
\vspace*{-0,3cm}
\end{figure*}

\vspace*{0,3cm}

\begin{figure*}[!t]
\begin{center}
  \includegraphics[scale=0.15]{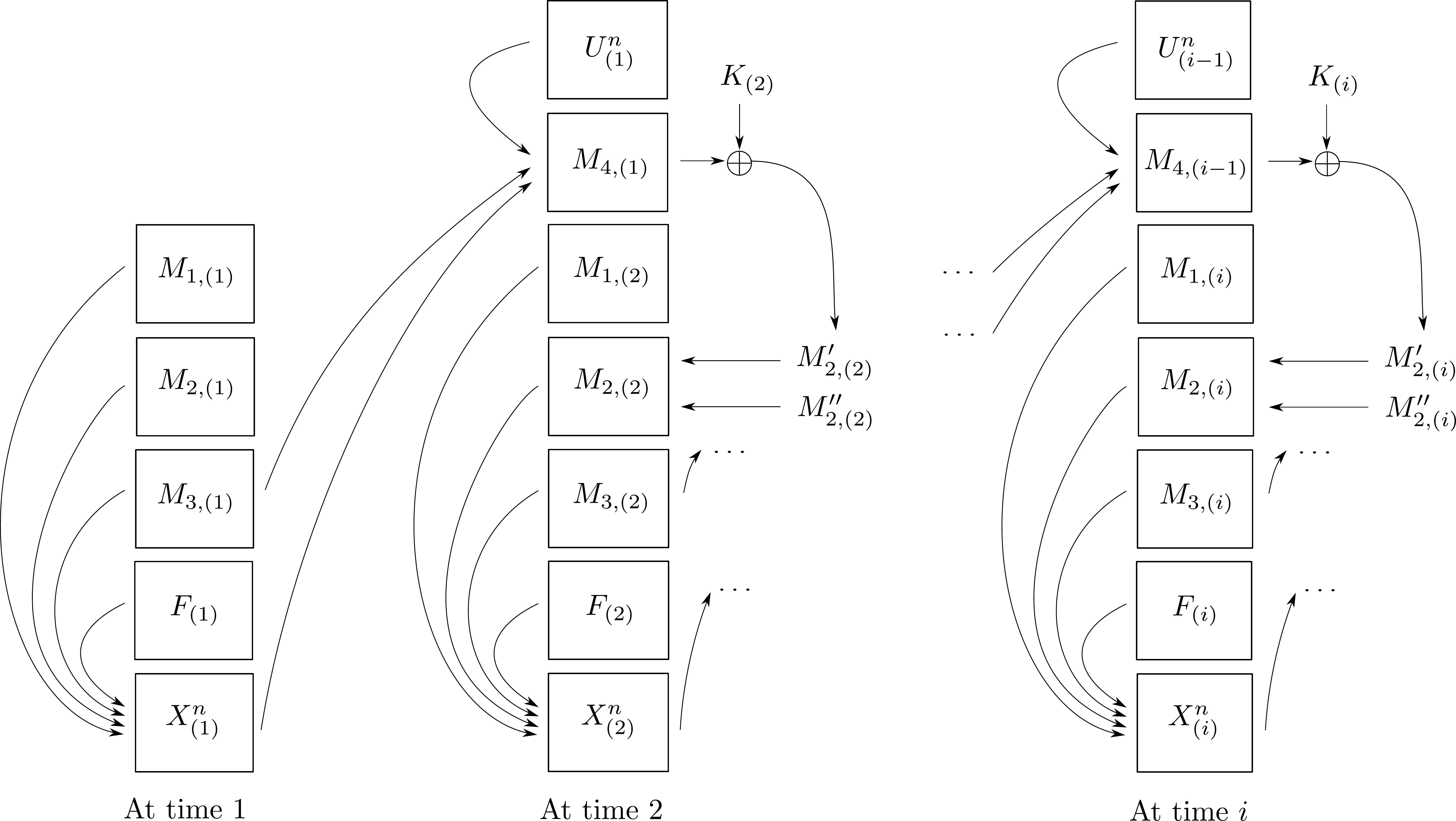}
\caption{Chaining construction for block Markov encoding}
\end{center}
\hrulefill
\label{fig: schemesc}
\end{figure*}

At time $i=1, \ldots, k$, the encoder does the following:
\begin{itemize}
\item  For $i \in \llbracket 1, k\rrbracket$, $M_{3,(i)}$ and $F_{(i)}$  are generated independently and uniformly  over  
$\llbracket 1, 2^{nR_3}\rrbracket$ and $\llbracket 1, 2^{n R}\rrbracket$ using common randomness with distributions $Q_{M_3}$ and $Q_{F}$ respectively; 
\item $M_{1,(i)}$  is generated independently and uniformly 
 over  $\llbracket 1, 2^{nR_1}\rrbracket$ using common randomness with distribution $Q_{M_{1}}$; 
\item In the first block, $M_{2,(1)}$ is generated uniformly at random using some independent local randomness;
\item For $i \in \llbracket 2, k\rrbracket$,  $M_{4,(i-1)}$ is generated according to the distribution defined earlier
 {\allowdisplaybreaks
 \begin{align*}
  &P^{\text{RB}}_{M_4|M_3 X^n U^n}(\mathbf m_{4,(i-1)}| \mathbf m_{3,(i-1)},  \mathbf x_{(i-1)}, \mathbf u_{(i-1)});
 \end{align*}}where  $(\mathbf m_{3,(i-1)},\mathbf x_{(i-1)}, \mathbf u_{(i-1)})$ are generated at time $i-1$;
\item For $i \in \llbracket 2, k\rrbracket$, $M_{2,(i)}= (M'_{2,(i)}, M''_{2,(i)})$, where 
\begin{equation}\label{onetimepad}
 M'_{2,(i)}= M_{4,(i-1)} \oplus K_i
\end{equation}
and  $K_i$ is generated uniformly over $\llbracket 1, 2^{n R_4}\rrbracket$ using common randomness,
 while $M''_{2,(i)}$ is generated uniformly at random using some independent local randomness.
 Thanks to the Crypto Lemma  \cite[Lemma 3.1]{bloch2011physical}, the distribution on $M_{2,(i)}$ is uniform and we denote it with $Q_{M_2}$;
\item The encoder generates $X^n_{(i)}$ according to the distribution defined earlier
 {\allowdisplaybreaks
 \begin{align*}
  &  P^{\text{RB}}_{X^n|M_1 M_2 M_3 F} (\mathbf x_{(i)}| \mathbf m_{1,(i)}, \mathbf m_{2,(i)}, \mathbf m_{3,(i)}, \mathbf f_{(i)});
 \end{align*}}Note that this distribution satisfies the strictly causal constraint, since $X^n_{(i)}$ is generated knowing the common randomness and 
 $M'_{2,(i)} =M_{4,(i-1)}\oplus K_i$, where $M_{4,(i-1)}$ depends on the source at time $i-1$;
\end{itemize}Then, the sequence $X^n_{(i)}$ is sent through the channel. 

\begin{oss}
Observe that we have imposed the condition $\lvert M_{4,(i-1)} \rvert= \lvert M'_{2,(i)}\rvert$, which holds as long as $R_4 \leq R_2$.
We have
{\allowdisplaybreaks
\begin{align*}
  R_2 &\!< \!H(X)\!- \!R_1<\! H(X)\!-\!H(X|Y)
   =I(X;Y),\\
 R_4&\!>I(W;U|X).
\end{align*}}Then, $R_4 \leq R_2$ implies $I(W;U|X)< I(X;Y)$.
\end{oss}

\subsubsection{Decoder}
Since the decoder is non-causal, it observes $Y^n_{(1:k)}$ and common randomness
$(M_{1,(1:k)}, M_{3,(1:k)}, F_{(1:k)}, K_{(2:k)})$ and the decoding algorithm proceeds as follows:

\begin{itemize}
\item The decoder reconstructs $\hat X^n_{(1:k)}$, where, for all $i \in \llbracket 1,k \rrbracket$, $\hat X^n_{(i)}$ is generated via the conditional distributions   
 {\allowdisplaybreaks
 \begin{align*}
P^{\text{SW}}_{\hat X^n |M_1  Y^n}(\mathbf x_{(i)}| \mathbf m_{1,(i)}, \mathbf y_{(i)});
 \end{align*}}
\item The decoder recovers $\hat M_{2,(1:k)}$, where, for all $i \in \llbracket 1,k\rrbracket$, $\hat M_{2,(i)}$ is generated 
via 
 \begin{equation*}
  \varphi_2(  \hat{\mathbf x}_{(i)})= \mathbf m_{2,(i)};
 \end{equation*}where  $\hat{\mathbf x}_{(i)}$ is the output of the Slepian-Wolf decoder;
 \item For all $i \in \llbracket 2,k \rrbracket$, with the key of the one-time pad $K_{(i)}$ and $\hat M'_{2,(i)}$,
 the decoder recovers $$\hat M_{4,(i-1)}=\hat M'_{2,(i)} \oplus K_{(i)};$$
\item Observe that at time $i$, the decoder knows an estimate of $\hat M_{4,(i)}$ because the non-causal nature of the decoder allows us to decode in reverse order and we note 
its distribution $P^{\text{RC}}_{\hat M_4}(\hat{\mathbf m}_{4,(i)})$. Therefore, once the decoder has $\hat M_{4,(i)}$, it reconstructs $ W^n_{(i)}$,
$i \in \llbracket 1,k-1 \rrbracket$, via
 {\allowdisplaybreaks
 \begin{align*}
  &P^{\text{SW}}_{W^n |M_3 \hat M_4 F \hat X^n}({\mathbf w}_{(i)}| \mathbf m_{3,(i)},\hat{\mathbf m}_{4,(i)}, \mathbf f_{(i)}, \hat{\mathbf x}_{(i)});
 \end{align*}}
\item Finally, the decoder generates $V^n_{(i)}$, $i \in \llbracket 1,k-1 \rrbracket$, 
letter by letter according to the distribution 
 {\allowdisplaybreaks
 \begin{align*}
P^{\text{RC}}_{V^n|  W^n  Y^n}({\mathbf v_{(i)}}|{\mathbf w}_{(i)},\mathbf y_{(i)}).
 \end{align*}} 
\end{itemize}For all $i \in \llbracket 1,k-1 \rrbracket$, the block-Markov coding scheme defines the joint distribution 
$P^{\text{RC}}_{(i)}$ in \eqref{rci}. 
\addtocounter{equation}{1}

\begin{oss}
Observe that, even though the block-Markov algorithm is over $k$ blocks,
the last block is only used to convey information on the source at time $k-1$ through $M_{4, (k-1)}$ which is generated at time $k$.
In fact, if $k$ is large enough, Definition \ref{definition sc ccord} allows us to coordinate only the first $k-1$ blocks.
\end{oss}

Now, observe that we impose rate conditions $R_1>H(X|Y)$ 
such that $\mathbb P \{\hat X_{(i)}^n \neq X_{(i)}^n\} \leq \delta(n)$ which in turn implies 
$\mathbb P \{\hat M_{2,(i)} \neq M_{2,(i)}\} \leq \delta(n)$, $\mathbb P \{\hat M_{4,(i)} \neq M_{4,(i)}\} \leq \delta(n)$. 
Moreover, 
{\allowdisplaybreaks
\begin{align*}
& \mathbb P \{\hat X_{(1:k)}^n \neq X_{(1:k)}^n\}  \leq  \sum_{i=1}^k \mathbb P \{\hat X_{(i)}^n \neq X_{(i)}^n\}  \leq k \delta(n),\\
& \mathbb P \{\hat M_{2,(1:k)} \neq M_{2,(1:k)}\} \leq \sum_{i=1}^k \mathbb P \{\hat M_{2,(i)} \neq M_{2,(i)}\} \leq k \delta(n), \\
&\mathbb P \{\hat M_{4,(1:k-1)} \neq M_{4,(1:k-1)}\} \leq \sum_{i=1}^{k-1} \mathbb P \{\hat M_{4,(i)} \neq M_{4,(i)}\} \stepcounter{equation}\tag{\theequation}\label{sw unionb}\\
&\phantom{\mathbb P \{\hat M_{4,(1:k-1)} \neq M_{4,(1:k-1)}\} } \leq (k-1) \delta(n),
\end{align*}}where $k \delta(n)$ and $(k-1) \delta(n)$ vanish since $\delta(n)$ goes to zero exponentially fast. 

We recall the definition of coupling and the basic coupling inequality for two random variables \cite{Lindvall1992coupling}.
\begin{defi}\label{defcoup}
A coupling of two probability distributions $P_A$ and $P_{A'}$ on the same measurable space $\mathcal A$ is any probability distribution 
$\hat P_{AA'}$ on the product measurable space $\mathcal{A} \times \mathcal{A}$  whose marginals are $P_A$ and $P_{A'}$.
\end{defi}

\begin{prop}[$\mbox{\cite[I.2.6]{Lindvall1992coupling}}$]\label{theocoup}
Given two random variables $A$, $A'$ with probability distributions $P_{A}$, $P_{A'}$, any coupling 
$\hat P_{AA'}$ of $P_{A}$, $P_{A'}$ satisfies
\begin{equation*}
\mathbb V(P_A, P_{A'})\leq 2 \mathbb P_{\hat P_{AA'}}\{A \neq A'\}.
\end{equation*}
\end{prop}
Then, we apply Proposition \ref{theocoup} to 
{\allowdisplaybreaks
\begin{align*}
A = &{(U^n  X^n Y^n V^n W^n  M_{1} M_{2} M_{3} M_{4} F )}_{(i)},\\
 A'=&{(U^n  \hat X^n Y^n V^n W^n  M_{1} M_{2} M_{3} \hat M_{4} F )}_{(i)},\\
 P_{A} =&P^{\text{RC}}_{A} \quad P_{A'}=P^{\text{RC}}_{A'} \\
  \mathcal A=& \mathcal U \times \mathcal X  \times \mathcal W  \times \mathcal Y \times  \llbracket 1,2^{nR_1} \rrbracket \times  \llbracket 1,2^{n R_2} \rrbracket \\
  & \times  \llbracket 1,2^{n R_3} \rrbracket \times \llbracket 1,2^{n R_4} \rrbracket \times\llbracket 1,2^{n \tilde R} \rrbracket. 
\end{align*}}and because of \eqref{sw unionb}
the distribution $\hat P^{\text{RC}}_{(i)}$  defined in \eqref{rci tilde} 
has almost the same statistics of $P^{\text{RC}}_{(i)}$:
\begin{equation*}
 \mathbb V (P^{\text{RC}}_{(i)}, \hat P^{\text{RC}}_{(i)}) \leq \delta(n).
\end{equation*}
\addtocounter{equation}{1}

\subsection{Coordination of ${(U^n, X^n, W^n, Y^n, V^n)}_{(i)}$}\label{conv1block}

 We want to show that the distribution $\hat P^{\text{RC}}_{(i)}$ is achievable  for strong coordination, i.e.,
\begin{equation}\label{shat}
\lim_{n \to \infty} \mathbb V \left( P^{\text{RB}}, \hat P^{\text{RC}}_{(i)}\right)=0.
\end{equation}

Observe that
\begin{itemize}
\item  By Lemma \ref{cuff17}  the total variational distance remains the same without $\bar P_{Y^n|X^n}$ and   $P_{V^n| W^n   Y^n}$  
in both $P^{\text{RB}}$ and $\hat P^{\text{RC}}_{(i)}$;
 \item The random binning distribution becomes
 \begin{align*}
 P^{\text{RB}}_{M_1 M_2 M_3 F X^n U^n}  P^{\text{RB}}_{M_4|M_3 X^n U^n} P^{\text{RB}}_{W^n|M_3 M_4 F X^n}
 \end{align*}and $ P^{\text{RB}}_{X^n|M_1 M_2 M_3 F} P^{\text{RB}}_{M_4|M_3 X^n U^n} P^{\text{RB}}_{W^n|M_3 M_4 F X^n} $ 
 can be removed in both $P^{\text{RB}}$ and $\hat P^{\text{RC}}_{(i)}$ by Lemma \ref{cuff17};
 \item Now, \eqref{shat} is satisfied if 
 \begin{align*}
 &\mathbb V \left( P^{\text{RB}}_{M_1 M_2 M_3 F  U^n} , \hat P^{\text{RC}}_{(M_1 M_2 M_3 F  U^n)_{(i)}}\right)\\
 &= \mathbb V \left( P^{\text{RB}}_{M_1 M_2 M_3 F  U^n} , Q_{M_3} Q_{F} Q_{M_1} Q_{M_2} P^{\text{RB}}_{ U^n} \right)
\end{align*}vanishes. By Lemma \ref{1.4.2}, this would be true if 
\begin{equation}\label{rate cond 1}
 R_1+R_2+R_3+\tilde R<H(WX|U)
\end{equation}.Since we have imposed the rate condition $R_3+\tilde R<H(W|XU)$ and $R_1+R_2<H(X)$ and $H(W|XU)+H(X)=H(WX|U)$ because
$X$ and $U$ are independent, \eqref{rate cond 1} holds and there exists a binning of $(W,X)$ such that
 \begin{align*}
  \mathbb{V}(P^{\text{RB}}_{M_1 M_2 M_3 F  U^n} , Q_{M_3} Q_{F} Q_{M_1} Q_{M_2} P^{\text{RB}}_{ U^n}) \leq \delta(n).
 \end{align*}Then we conclude that \eqref{shat} holds.
\end{itemize}

\subsection{Coordination of ${(U^n, X^n,  Y^n, V^n)}_{(i)}$ by removing the extra randomness F}\label{rf gen}

Even though the extra common randomness $F$ is required to coordinate $\left(U^{n}\right.$,  ${X}^{n}$, $Y^{n}$, $V^{n}$, $\left.W^{n}\right)$ 
we will show that we do not need it in order to coordinate only $(U^{n},{X}^{n},Y^{n},V^{n})$.
As in \cite{yassaee2014achievability}, we would like to reduce the amount of common randomness by having the two nodes agree on an instance $F=f$.
To do so, we apply Lemma \ref{1.4.2} again where $B^{n}=W^{n}$,  
$K=F$, $\varphi$ and $A^{n}= U^{n} X^{n} Y^{n} V^{n}$.
If $\tilde R < H(W| UXYV)$, there exists a fixed binning such that
 {\allowdisplaybreaks
\begin{align*}
\mathbb V (P^{\text{RB}}_{U^{n} X^{n} Y^{n} V^{n} F}, Q_F P^{\text{RB}}_{U^{n} X^{n} Y^{n} V^{n}})=\delta(n). \stepcounter{equation}\tag{\theequation}\label{bin3}  
\end{align*}}which implies
 {\allowdisplaybreaks
\begin{align*}
 &  \mathbb V (\hat P^{\text{RC}}_{ {(U^n X^n Y^n V^n F)}_{(i)}}, Q_F P^{\text{RB}}_{U^{n} X^{n} Y^{n} V^{n}})=\delta(n). 
 \stepcounter{equation}\tag{\theequation}\label{bin4}  
\end{align*}}By  Lemma \ref{lem4}, there exists an instance $f \in \llbracket 1, 2^{n \tilde R}\rrbracket$ such that
 {\allowdisplaybreaks
\begin{align*}
& \mathbb V (P^{\text{RB}}_{ U^{n} X^{n} Y^{n} V^{n}|F=f},\hat P^{\text{RC}}_{ {(U^n X^n Y^n V^n)}_{(i)} |F_{(i)}=f}) =\delta(n).
\stepcounter{equation}\tag{\theequation}\label{condf}\end{align*}}Then, by fixing $F=f$ and using common randomness 
$C=(M_{1,(i)}, M_{3,(i)}, K_{i})$, we have coordination for $(U^n, X^n, Y^n, V^n)$.

\subsection{Rate of common randomness}
We have used common randomness to generate $M_1$,$M_3$ and the key of the one time pad, which has the same size of $M_4$.
Then, upon denoting by $R_0$ the total rate of common randomness, $R_0:=R_1+R_3+R_4$ and 
\begin{align*}
& R_0+ \tilde R >  H(X|Y)+H(W|X)\\
&  \tilde R < H(W| UXYV)
\end{align*}
which implies
\begin{equation}\label{rate cr}
  R_0> H(X|Y)+H(W|X)-H(W|UXYV).
\end{equation}Observe that 
{\allowdisplaybreaks
\begin{align*}
 H(WX|Y)&=H(WX)-I(WX;Y)\\
 &=H(X)+H(W|X)-I(X;Y)\\
 &=H(X|Y)+H(W|X)
\end{align*}}because the Markov chain $W-X-Y$ implies $I(W;Y|X)=0$ and therefore \eqref{rate cr} becomes
{\allowdisplaybreaks
\begin{align*}
 R_0&> H(WX|Y)-H(W|UXYV)\\
 &=H(W|Y)+H(X|WY)-H(W|UXYV) \stepcounter{equation}\tag{\theequation}\label{rate cr2}\\
 &=I(W; UXV|Y)+H(X|WY).
\end{align*}}

\subsection{Coordination of all blocks}\label{convallblocks rb}

To simplify the notation, we set 
 {\allowdisplaybreaks
 \begin{equation*}
\begin{matrix}[ll]
  L_{i}:=U^{n}_{(i)} X^{n}_{(i)} Y^{n}_{(i)} V_{(i)}^n & i \in \llbracket 1,k-1\rrbracket\\
  L_{a:b}:=U^{n}_{(a:b)} X^{n}_{(a:b)}  Y^{n}_{(a:b)} V^{n}_{(a:b)} & \llbracket a,b\rrbracket \subset \llbracket 1,k-1\rrbracket.
\end{matrix}
\end{equation*}
}

First, note that two consecutive blocks
$L_{i-1}$ and $L_{i}$  are dependent only through $M_{4,(i-1)}$.
In fact, $M_{4,(i-1)}$ is created at time $i$ using $U^{n}_{(i-1)}$ and $X^{n}_{(i-1)}$ 
and it is used to generate $M_{2,(i)}$, which in 
turn is used at the encoder to generate $X^{n}_{(i)}$.
Hence, since $Y^{n}_{(i)}$ is the output of the channel and $V^{n}_{(i)}$ is 
generated using $Y^{n}_{(i)}$ and the auxiliary random variable, generated through an estimate of $\hat M_{4,(i)}$, 
uniform common randomness and  $X^{n}_{(i)}$, we can conclude that $L_{i-1}$ and $L_{i}$ 
are dependent only through $M'_{2,(i)}$ and therefore $M_{4,(i-1)}$.
However, to generate $M_{2,(i)}$, the encoder applies a one-time pad on $M_{4,(i-1)}$ as shown in~\eqref{onetimepad}, 
making $M_{4,(i-1)}$ and $M_{2,(i)}$ independent of each other and ensuring the independence of two consecutive blocks.

To conclude the proof we need the following results.

\begin{lem}\label{step2b'rb}
We have
$$ \mathbb V \left( P_{ L_{1:k-1}}, \prod_{i=1}^{k-1}  P_{ L_{i}} \right) \leq  \delta(n).$$
\end{lem}

\begin{lem}\label{step2c'rb}
We have
$$ \mathbb V \left( P_{L_{1:k-1}}, \bar P_{UXYV}^{\otimes n(k-1)}\right) \leq \delta(n).$$
\end{lem} 

We omit the proofs because they are very similar to the proofs of \cite[Lemma 15]{Cervia2018journal} and \cite[Lemma 16]{Cervia2018journal} respectively.

\section{Outerbound of Theorem \ref{teostrictly}}\label{outerpart1}
Consider a code $(f^n,g^n)$ that induces a distribution $P_{U^{n}  X^n Y^n V^{n}}$ 
that is $\varepsilon$-close in total variational distance to the i.i.d. distribution $\bar P_{UXY V}^{\otimes n}$.
Let the random variable $T$ be uniformly distributed over the
set $\llbracket 1,n\rrbracket$ and independent of the sequence
$(U^{n}, X^{n}, Y^{n}, V^{n}, C)$. The variable $T$ will serve as a random time index. 
The variable $U_T$ is independent of $T$ because $U^{n}$ is an i.i.d. source \cite{cuff2010}.

\subsection{Bound on $R_0$}\label{convpart1}

We have
{\allowdisplaybreaks
\begin{align*}
& nR_0 = H(C) \geq H(C|Y^n) \geq I(C;U^{n} V^n X^n|Y^n)\\
& = \sum_{t=1}^n I(U_t V_t X_t ;C| U^{t-1} V^{t-1} X^{t-1} Y_{\sim t} Y_{t}) \\
& = \sum_{t=1}^n I(U_t V_t X_t ;C Y_{\sim t}  U^{t-1} V^{t-1} X^{t-1}| Y_t)\\
& \phantom{=} -  \sum_{t=1}^n I (U_tV_t X_t ;Y_{\sim t}  U^{t-1} V^{t-1} X^{t-1}| Y_t)\\
& \overset{(a)}{\geq} \sum_{t=1}^n I(U_t V_t X_t ;C Y_{\sim t}  U^{t-1} V^{t-1} X^{t-1}| Y_t) -n f(\varepsilon)\\
&\geq \sum_{t=1}^n I(U_t V_t X_t ;C Y_{t+1}^n  U^{t-1} | Y_t) -n f(\varepsilon)\\
&  =\! n I(U_T V_T X_T;C Y_{T+1}^n U^{T-1} | Y_T T) -nf(\varepsilon)\\
& =\! n I(U_T V_T X_T;C Y_{T+1}^n U^{T-1} T| Y_T )\\
& \phantom{=}-n I(U_T V_T X_T;T|Y_T) -nf(\varepsilon)\\
& \geq  n I(U_T V_T X_T;C Y_{T+1}^n U^{T-1} T| Y_T ) \\
& \phantom{=}-n I(U_T V_T X_T Y_T;T) -nf(\varepsilon)\\
&\overset{(b)}{\geq} n I(U_T V_T X_T;C Y_{T+1}^n U^{T-1} T| Y_T ) - 2nf(\varepsilon)
\end{align*}}where $(a)$ comes from Lemma \ref{lemmit} and $(b)$ comes from \cite[Lemma VI.3]{cuff2013distributed}.

\subsection{Information constraint}\label{convpart2}
We have
{\allowdisplaybreaks
\begin{align*}
&  n I(U_T;C Y_{T+1}^n U^{T-1} X_T T)\\
&\overset{(a)}{=} n I(U_T;C Y_{T+1}^n U^{T-1} X_T |T) 
\! = \! \sum_{t=1}^n I(U_t;C Y_{t+1}^n U^{t-1} X_t)\\
& = \!\sum_{t=1}^n I(U_t;C Y_{t+1}^n U^{t-1} ) \! + \! \sum_{t=1}^n I(U_t;X_t|C Y_{t+1}^n U^{t-1})\\
& \leq  \!\sum_{t=1}^n I(U_t;C Y_{t+1}^n U^{t-1} )  \!+  \!\sum_{t=1}^n I(U_t Y_{t+1}^n ;X_t|C  U^{t-1})\\
&\overset{(b)}{=} \sum_{t=1}^n I(U_t;C Y_{t+1}^n U^{t-1} ) \overset{(c)}{=} \! \sum_{t=1}^n I(U_t; Y_{t+1}^n | U^{t-1} C) \\
& \overset{(d)}{=} \sum_{t=1}^n I(Y_t; U^{t-1}| Y_{t+1}^n C) 
\leq \sum_{t=1}^n I(Y_t; U^{t-1} Y_{t+1}^n C) \\
& \leq \sum_{t=1}^n I(Y_t; U^{t-1} Y_{t+1}^n C X_t) = n I(Y_T; U^{T-1} Y_{T+1}^n C X_T|T) \\
&\leq n I(Y_T; U^{T-1} Y_{T+1}^n C X_T T) 
\end{align*}}where $(a)$ follows from the i.i.d. nature of the source, $(b)$ from the following Markov chain
\begin{equation*}
 X_t-(C,  U^{t-1})-(U_t, Y_{t+1}^n)
\end{equation*}that holds because of the strictly causal nature of the encoder.
Then, $(c)$ comes from the fact that the source is generated i.i.d. and independent of $C$ and 
$(d)$ from Csisz{\'a}r's sum identity. 

We identify the auxiliary random variables $W_t$ with $(U^{t-1}, Y_{t+1}^n, C)$ for each $t \in \llbracket 1,n\rrbracket$ and $W$ with
$(W_T,T)=(U^{T-1}, Y_{T+1}^n, C, T)$. 

\subsection{Identification of the auxiliary random variable}\label{convpart3}
For each $t \in \llbracket 1,n\rrbracket$, $W_t$ satisfies the following conditions:
\begin{align*}
 & U_t \perp  X_t \\
 & Y_t - X_t- (U_t, W_t) \stepcounter{equation}\tag{\theequation}\label{channelt}\\
 & V_t - (Y_t, W_t)-(U_t, X_t).
\end{align*}Then, we have 
\begin{align*}
 & U_T \perp  X_T\\
 & Y_T - X_T- (U_T, W_T) \stepcounter{equation}\tag{\theequation}\label{channelT}\\
 & V_T - (Y_T, W_T)-(U_T, X_T),
\end{align*}and, since $W=W_t$ when $T=t$, it implies
\begin{align*}
 & U \perp  X\\
 & Y - X- (U, W)\stepcounter{equation}\tag{\theequation} \label{channel}\\
 & V - (Y, W)-(U, X).
\end{align*}We do not write all details because they follow similarly the discussion in \cite[Section VIII-B]{treusttech}.
The proof of the cardinality bound is omitted since it follows the ones in \cite[Appendix G]{Cervia2018journal}.

\appendix[Explicit polar coding scheme] \label{inner polar}

In this section, we propose a polar coding scheme that achieves the region $\mathcal R_{\text{in}}$.
For brevity, we only focus on the set of achievable distributions in $\mathcal R_{\text{in}}$ for which the auxiliary variable $W$ is binary. 
The scheme can be extended to the case of a non-binary random variable $W$ using non-binary polar codes as long as the cardinality 
$\lvert \mathcal W\rvert$ is a prime number \cite{csacsouglu2012polar}.
\subsection{Polar coding scheme}\label{sec:polar}
Assume that the sequences $U^n$, $X^n$, $W^n$, $Y^n$ and $V^n$ are jointly i.i.d. with distribution \eqref{iiddistr}.
We propose an explicit coding scheme similar to the one in \cite{Cervia2017gretsi} that induces a joint distribution close  to  \eqref{iiddistr} in total variational distance.

\vspace{0,3cm}
\paragraph{Polarize $X$}
Let $S^{n}=X^{n}G_n$ be the polarization of $X^{n}$, where $G_n$ is the source polarization transform.
For some $\mbox{$0<\beta<1/2$}$, let $\delta_n: = 2^{-n^ {\beta}}$ and define the very high and high entropy sets:
{\allowdisplaybreaks
 \begin{align*}
\mathcal V_{X}: & =\left\{j\in\llbracket 1,n \rrbracket \mbox{ } | \mbox{ } H(S_j|S^{j-1})>1-\delta_n \right\},\\
\mathcal H_{X}: & =\left\{j\in\llbracket 1,n \rrbracket \mbox{ } | \mbox{ } H(S_j|S^{j-1})>\delta_n \right\},  \stepcounter{equation}\tag{\theequation}\label{eq: hz}\\
\mathcal H_{X | Y}: & =\left\{j\in\llbracket 1,n \rrbracket \mbox{ } | \mbox{ } H(S_j|S^{j-1} Y^{n})>\delta_n \right\} .
 \end{align*}}
Partition the set $\llbracket 1,n \rrbracket$ into four disjoint sets:
\begin{equation*}
\begin{matrix}[ll]
A_1 := \mathcal V_{X} \cap \mathcal H_{X|Y} , \quad & A_2 : = \mathcal V_{X} \cap \mathcal H_{X|Y}^c, \\
A_3 := \mathcal V_{X}^c \cap \mathcal H_{X|Y} ,\quad & A_4 := \mathcal V_{X}^c \cap \mathcal H_{X|Y}^c.
\end{matrix}
\end{equation*}

\begin{oss}\label{oss card1}
We have:
\begin{itemize}
\item[\textbullet] $\mathcal V_{X} \subset \mathcal H_{X}$ and $\displaystyle \lim_{n \rightarrow \infty} \lvert \mathcal H_{X} \setminus \mathcal V_{X}  \rvert/n = 0$, \cite{arikan2010source},
 \item[\textbullet] $A_1 \cup A_2 = \mathcal V_X$ and $\displaystyle \lim_{n \rightarrow \infty} \lvert \mathcal V_{X} \rvert/n  = H(X)$  \cite {chou2015secretkey},
 \item[\textbullet]  $A_1 \cup A_3 = \mathcal H_{X|Y}$ and $\displaystyle \lim_{n \rightarrow \infty} \lvert \mathcal H_{X | Y} \rvert/n = H(X|Y)$ \cite{arikan2010source}.
\end{itemize}
Since  $\displaystyle \lim_{n \rightarrow \infty} \frac{\lvert A_2\rvert - \lvert A_3 \rvert}{n}= H(X)- H(X|Y) = I(X;Y)\geq 0$ 
this implies directly that for $n$ large enough $\lvert A_2 \rvert \geq \lvert A_3 \rvert$.
\end{oss}

\vspace{0,3cm}
\paragraph{Polarize $W$}
Let $Z^{n}=W^{n}G_n$ be the polarization of $W^{n}$ and define:
{\allowdisplaybreaks
 \begin{align*}
\mathcal V_{W|XU}: & =\left\{j\in\llbracket 1,n \rrbracket \mbox{ } | \mbox{ } H(Z_j|Z^{j-1}X^{n}U^{n})>1-\delta_n \right\},\\
\mathcal H_{W | XU}: & =\left\{j\in\llbracket 1,n \rrbracket \mbox{ } | \mbox{ } H(Z_j|Z^{j-1} X^{n}U^{n})>\delta_n \right\}, \stepcounter{equation}\tag{\theequation}\label{eq: hv} \\
\mathcal H_{W | X}: & =\left\{j\in\llbracket 1,n \rrbracket \mbox{ } | \mbox{ } H(Z_j|Z^{j-1} X^{n})>\delta_n \right\} .
 \end{align*}}

Partition the set $\llbracket 1,n \rrbracket$ into four disjoint sets:
\begin{equation*}
\begin{matrix}[l]
B_1 := \mathcal V_{W|XU} \cap \mathcal H_{W|X}=\mathcal V_{W|XU} ,\\
B_2 : = \mathcal V_{W|XU} \cap \mathcal H_{W|X}^c=\emptyset,\\
B_3 := \mathcal V_{W|XU}^c \cap \mathcal H_{W|X} ,\\
B_4 := \mathcal V_{W|XU}^c \cap \mathcal H_{W|X}^c=\mathcal H_{W|X}^c.
\end{matrix}
\end{equation*}

\begin{oss}\label{oss card2}
We have:
\begin{itemize}
 \item[\textbullet] $\!\!\mathcal V_{W|XU}\!\!\subset\! \mathcal H_{W|XU}$ and $\displaystyle\! \lim_{n \rightarrow \infty} \lvert \mathcal H_{W|XU}\! \setminus\! \mathcal V_{W|XU}  \rvert/n\!\! =\!\! 0$ \cite{arikan2010source},
 \item[\textbullet] $\!\!B_1  = \mathcal V_{W|XU}$ and $\displaystyle \lim_{n \rightarrow \infty} \lvert\mathcal V_{W|XU}\rvert/n  = H(W|XU)$ \cite {chou2015secretkey},
  \item[\textbullet]  $\!\!B_4 = \mathcal H_{W|X}^c$ and $\displaystyle \lim_{n \rightarrow \infty} \lvert \mathcal V_{W|X}^c \rvert/n = 1 - H(W|X)$ \cite {chou2015secretkey},
 \item[\textbullet]  $\!\!B_3\cup B_4 \!=\! \mathcal V_{W|XU}^c$ and 
 $ \displaystyle \lim_{n \rightarrow \infty} \lvert \mathcal V_{W|XU}^c \rvert/n =\! 1 - \!H(W|XU)$\cite{chou2015secretkey}.
\end{itemize}
Note that  $$H(W|X)- H(W|XU) = I(W;U|X) =  I(WX;U)\geq 0$$ and
 $\lvert B_3 \rvert/n $ tends to $I(WX;U)$. Since $I(WX;Y) = I(X;Y)$, the inequality $I(WX;U) \leq I(WX;Y)$  implies directly that for $n$ large enough 
 $\lvert B_3 \rvert \leq \lvert A_2\rvert - \lvert A_3 \rvert $.
\end{oss}

\begin{algorithm}[h!]\label{alg1}
\DontPrintSemicolon
\SetAlgoVlined 
\SetKwInOut{Input}{Input}
\SetKwInOut{Output}{Output}
\Input{ $(U_{(0)}^{n}, \ldots, U_{(k)}^{n})$, local randomness (uniform random bits) $M$ and 
common randomness $C=({\{C_{i}\}}_{i=1, \ldots ,k}, {\{C'_{i}\}}_{i=1, \ldots ,k}, \bar C', {\{K_{i}\}}_{i=1, \ldots ,k} ,$ 
$ {\{K'_{i}\}}_{i=1, \ldots ,k})$ shared with Node 2:
\begin{itemize}\setlength{\itemsep}{0.2em}
\item [-] $C_i$ of size $\lvert A_1 \rvert$ and $K_i$ of size $\lvert A_3 \rvert$
\item [-] $\bar C'$ of size $\lvert B'_1 \rvert$, $C'_i$ of size $\lvert B_1 \setminus B'_1\rvert$ and $K'_i$ of size $\lvert B_3 \rvert$
\end{itemize}}

\Output{$(S^{n}_{(1)}, \ldots, S^{n}_{(k)} )$, $(Z^{n}_{(1)}, \ldots,Z^{n}_{(k)})$}
\If{$i=1$}{
 $S_{(1)}[A_1] \longleftarrow C_i, \mbox{ } S_{(1)}[A_2] \! \longleftarrow M $\;
\For{$j \in A_{3} \cup A_{4}$}{Successively draw the bits $S_{j,(1)}$ according to 
\vspace{-0.2cm}
\begin{small}
\begin{equation} \label{eq: p11}
\bar P_{S_j | S^{j-1}} (S_{(i),j}| S_{(i)}^{j-1} )
\end{equation}
\end{small}
\vspace{-0.5cm}}\;
\vspace{-0.4cm}
$Z_{(1)}[B'_1] \longleftarrow \bar C' \quad Z_{(1)}[B_1 \setminus B'_1] \longleftarrow C'_i$\;
\For{$j \in B_{3} \cup B_{4}$}{
Given $U_{(1)}^{n}$, successively draw the bits $Z_{(1)}^j$ according to 
\vspace{-0.2cm}
\begin{small}
\begin{equation} \label{eq: p21}
\bar P_{Z_j | Z^{j-1}X^{n}U^{n}} (Z_{(i),j}| Z_{(i)}^{j-1} X_{(i)}^n U_{(i-1)}^n )
\end{equation}
\end{small}
\vspace{-0.5cm}}}

\For{$i=2, \ldots, k$}{
\vspace{-0.4cm}
\begin{align*}
 & S_{(i)}[A_1] \longleftarrow C_i, \quad S_{(i)}[A'_2] \longleftarrow M \\  
 & S_{(i)}[B'_3] \longleftarrow Z_{(i-1)}[B_3] \oplus K_{i-1} \\
 & S_{(i)}[A'_3] \longleftarrow S_{(i-1)}[A_{3}] \oplus K_{i-1}
 \end{align*}\;
 \vspace{-0.5cm}
\For{$j \in A_{3} \cup A_{4}$}{
Succ.  draw the bits $S_{(i),j}$ according to \eqref{eq: p11}\;
\vspace{-0.2cm}
\begin{align*}
 &Z_{(i)}[B'_1] \longleftarrow \bar C'\\
 &Z_{(i)}[B_1\setminus B'_1] \longleftarrow C'_{i} 
\end{align*}\;
\For{$j \in B_{3} \cup B_{4}$}{Succ. draw the bits $Z_{(i),j}$ according to \eqref{eq: p21}}}}
\BlankLine
\caption{Encoding algorithm at Node 1}
\end{algorithm}

\paragraph{Encoding}

The encoder observes  $U^{n}_{(0:k)}:=(U^{n}_{(0)}, U^{n}_{(1)}, \ldots, U^{n}_{(k)})$, where $U^{n}_{(0)}$ is a uniform random sequence and $U^{n}_{(i)}$
for $i \in \llbracket 1,k \rrbracket$ are $k$ blocks of the source.
It then generates for each block $i\in \llbracket 1,k\rrbracket$  random variables $S^{n}_{(i)}$  and $Z^{n}_{(i)}$ following the procedure described in Algorithm \ref{alg1}.
The chaining construction proceeds as follows:
\begin{itemize}
\item The bits in  $A_1 \subset \mathcal V_{X}$ in block $i \in \llbracket 1,k\rrbracket$ are chosen with uniform probability using a
uniform randomness source $C_{i}$ shared with the decoder;
\item In the first block the bits in  $A_2 \subset \mathcal V_{X}$ are chosen with uniform probability using a local randomness source $M$;

\item Let $B'_1:=\mathcal V_{W|UXYV}$, observe that $B'_1$ is a subset of $ B_1$ since
 $\mathcal V_{W|UXYV} \subset \mathcal V_{W|XU}$.
 The bits in  $B'_1 \subset \mathcal V_{W|XU}$ in block $i \in \llbracket 1,k\rrbracket$ are chosen with uniform probability using a uniform randomness 
 source $\bar C'$ shared with the decoder, and  their value is reused over all blocks;
\item The bits in  $B_1 \setminus B'_1 \subset \mathcal  V_{W|XU}$ in block $i \in \llbracket 1,k\rrbracket$ are chosen with uniform probability using a
uniform randomness source $C'_{i}$ shared with the decoder;

\item The bits in $A_3 \cup A_4$  and $B_3 \cup B_4$ are generated according to the previous bits using
successive cancellation encoding as in \cite{chou2015secretkey}.
Note that it is possible to sample efficiently from $\bar P_{S_j | S^{j-1}}$  and
$\bar P_{Z_j | Z^{j-1}X^{n}U^{n}}$ (given $U^{n}$ and $X^{n}$) respectively;

\item From the second block, the encoder generates the bits of $A_2$ in the following way. Let $A'_3$ and $B'_3$ be two disjoint 
subsets of $A_2$ such that $\lvert A'_3 \rvert= \lvert A_3 \rvert$ and $\lvert B'_3 \rvert= \lvert B_3 \rvert$.  
The existence of those disjoint subsets is guaranteed by Remark \ref{oss card1} and Remark \ref{oss card2}.
The bits of $A_3$ and $B_3$ in block $i$ are used as $A'_3$ and $B'_3$ in block $i+1$ using one-time pads with keys 
$K_i$ and $K'_i$ respectively:
{\allowdisplaybreaks
\begin{align*}
 & S_{(i+1)}[A'_3]=S_{(i)}[A'_3] \oplus K_i \quad  i = 1,\ldots,  k-1,\\
 & S_{(i+1)}[B'_3]=Z_{(i)}[A'_3] \oplus K'_i \quad  i=1,\ldots, k-1. 
\end{align*}
}
Thanks to the Crypto Lemma  \cite[Lemma 3.1]{bloch2011physical}, 
if we choose $K_i$ of size $\lvert A_3 \rvert$ and  $K'_i$ of size $\lvert B_3 \rvert$ 
to be uniform random keys, the bits in $A'_3$ and $B'_3$ in the block $i+1$ are uniform. 
The bits in $A'_2:=A_2 \setminus (A'_3 \cup B'_3)$ are chosen with uniform probability using the local randomness source $M$.

\end{itemize}

The encoder then computes 
$X_i^{n}=S_i^{n} G_n$  for $i=1, \ldots, k$ and sends it over the channel.
As in \cite{chou2015polar}, to deal with unaligned indices, chaining also requires in the last encoding block to transmit 
$S_{(k)}[A_{3}] \cup Z_{(k)}[B_{3}]$  to the decoder.
Hence the coding scheme requires an error-free channel between the encoder and decoder which has negligible rate 
since $\lvert S_{(k)}[A_{3}] \cup Z_{(k)}[B_{3}] \rvert \leq \lvert \mathcal H_{X} \rvert$ and
\begin{equation*}
\lim_{n \rightarrow \infty \atop k \rightarrow \infty} \frac{\lvert \mathcal H_{X}\rvert}{kn} = \lim_{k \rightarrow \infty} \frac{H(X)}{k} =0.
\end{equation*}

\begin{center}\label{fig:kblo}
\begin{figure}[h!]
 \centering
 \includegraphics[scale=0.72]{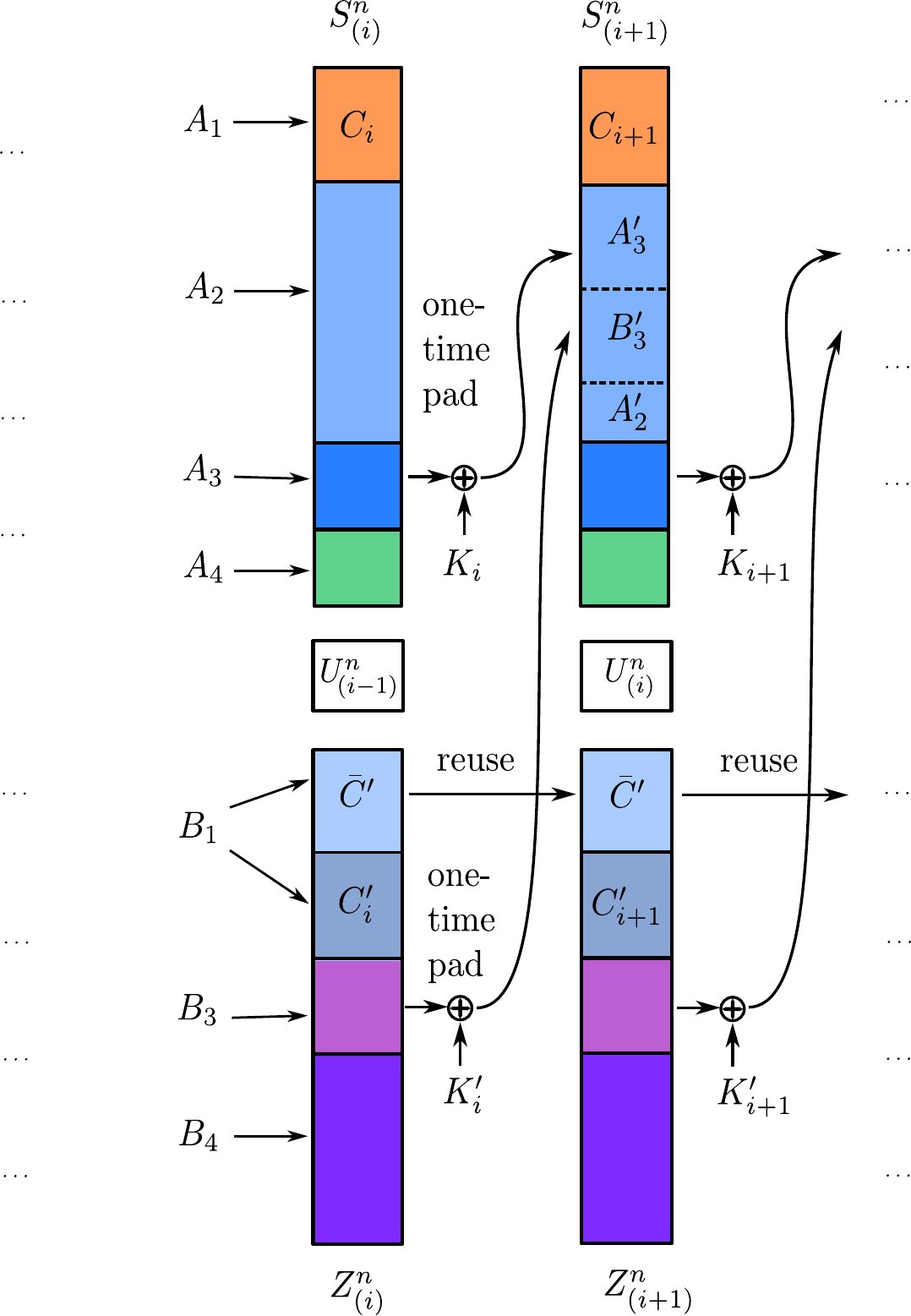}
\caption{Chaining construction for block Markov encoding with polar codes}
\end{figure}
\end{center}

\paragraph{Decoding} 
The decoder observes $(Y_{(1)}^{n}, \ldots, Y_{(k)}^{n})$ and  $S_{(k)}[A_{3}] \cup Z_{(k)}[B_{3}]$ allows it to decode in reverse order.
The decoding algorithm, detailed in Algorithm \ref{alg2}, proceeds as follows:
\begin{itemize}
 \item In every block $i \in \llbracket 1,k \rrbracket $,
 the decoder has access to $\hat S_{(i)}[A_{1}] $ $\subseteq \hat S_{(i)}[\mathcal H_{X| Y}]$ and 
$\hat Z_{(i)}[B_{1}]$ $\subseteq \hat Z_{(i)}[\mathcal H_{W |X}]$ because the bits in $A_1$ and $B_1$ correspond to shared randomness 
$({\{C_{i}\}}_{i=1, \ldots ,k},$ $ {\{C'_{i}\}}_{i=1, \ldots ,k}, \bar C')$,
\item In block $i \in \llbracket 1, k-1 \rrbracket$ the bits in $A_3$ and $B_3$ are obtained by successfully recovering $A_2$ in block $i+1$, which is
possible because the keys of the one-time pad are part of the common randomness;
\item From $Y_{(i)}^{n}$ and $\hat S_{(i)}[A_1 \cup A_{3}]$ the 
successive cancellation decoder  can retrieve $\hat S_{(i)}[A_2 \cup A_4]$ and  $\hat Z_{(i)}[B_4]$.
Note that, by \cite[Theorem 3]{arikan2010source}, $Z^{n}$ is equal to  $\hat Z^{n}$ and  $S^{n}$ is equal to  $\hat S^{n}$ with high probability.
\item The decoder computes $\hat W_{(i)}^{n} = \hat Z_{(i)}^{n} G_n $
\item Finally, the decoder generates $V_{(i)}^{n}$ symbol by symbol using $$P_{V_{(i),j} | \hat W{(i),j} Y_{(i),j}} (v|w,y)=\bar P_{V | WY}(v|w,y).$$
\end{itemize}

\vspace{-0.3cm}

\begin{algorithm}[h!]\label{alg2}
\DontPrintSemicolon
\SetAlgoVlined 
\SetKwInOut{Input}{Input}
\SetKwInOut{Output}{Output}
\Input{$(Y_{(1)}^{n}, \ldots, Y_{(k)}^{n})$, $S_{(k)}[A_{3}] \cup Z_{(k)}[B_{3}]$ and $C$ common randomness shared with Node 1}
\Output{$(\hat S_{(1)}^{n}, \ldots, \hat S_{(k)}^{n})$, $(\hat Z_{(1)}^{n}, \ldots, \hat Z_{(k)}^{n})$}
\For{$i=k, \ldots, 1$}{
\begin{align*}
 &  \hat S_{(i)}[A_1] \longleftarrow C_i \\
 & \hat Z_{(i)}[B'_1] \longleftarrow \bar C' \quad \hat Z_{(i)}[B_1 \! \setminus \! B'_1] \longleftarrow C'_i 
\end{align*}
 \If{$i\neq k$}{
 \begin{align*}
&  \hat S_{(i)}[A_3] \leftarrow \hat S_{(i+1)} [A'_{3}] \oplus K_{i}\\
& \hat Z_{(i)}[B_3] \leftarrow \hat S_{(i+1)} [B'_{3}] \oplus K'_{i}
 \end{align*}}
\For{$j \in A_{2} \cup A_{4}$}{ 
Successively draw the bits according to
\begin{align*}
 &\hat S_{(i),j} = \begin{cases}
 0 \quad \mbox{if } L_n(Y_{(i)}^{n}, \hat S_{(i)}^{j-1}) \geq 1\\
 1 \quad \mbox{else} 
 \end{cases}\\
 &\mbox{where }\\
 &L_n(Y_{(i)}^{n}, \hat S_{(i)}^{j-1}) =
 \frac{\bar P_{S_{j} | S^{j-1}  Y^{n}} \left(0 | \hat S_{(i)}^{j-1}  Y_{(i)}^{n} \right) }
 {\bar P_{S_{j} | S^{j-1}  Y^{n}} \left(1 | \hat S_{(i)}^{j-1}  Y_{(i)}^{n} \right)}
 \end{align*}

}
\For{$j \in B_{4}$}{ Successively draw the bits according to
$$\hat Z_{(i),j} = \begin{cases}
 0 \quad \mbox{if } L_n(X_{(i+1)}^{n}, \hat Z_{(i)}^{j-1}) \geq 1\\
 1 \quad \mbox{else} 
 \end{cases}$$
\vspace{-0.3cm}
}
}
\BlankLine
\caption{Decoding algorithm at Node 2}
\end{algorithm}

\paragraph{Rate of common randomness}
The rate of common randomness is $I(W;UXV|Y)+ H(X|WY)$ since:
{\allowdisplaybreaks
\begin{align*}
 &\lim_{n \rightarrow \infty \atop k \rightarrow \infty} \! \frac{ k \lvert A_1 \rvert\! +\! (k-1) \lvert A_3 \rvert\! +\! k \lvert B_1 \rvert \! + \!(k-1) \lvert B_3  \rvert\! - \!(k-1) \lvert B'_1 \rvert}{kn}\\
  &=\lim_{n \rightarrow \infty} \frac{  \lvert A_1 \rvert +  \lvert A_3 \rvert + \lvert B_1 \rvert  + \lvert B_3  \rvert -  \lvert B'_1 \rvert }{n}\\
 &= H(X|Y) + H(W|X)-H(W|UXYV)\\
 &{\overset{{(a)}}{=}} I(W;UXV|Y)+ H(X|YW)
\end{align*}}where $(a)$ has been proved in \eqref{rate cr2}.

\subsection{Coordination in one block}\label{conv1block polar}
We note with $P$ the joint distribution induced by the encoding and decoding algorithm of the previous sections. 
The proof requires a few steps. Similarly to \cite[Lemma 13]{Cervia2018journal}, we first prove that in each block $i \in \llbracket 1,k\rrbracket$ 
\begin{equation}\label{emp1}
 \mathbb D\left(  \bar P_{UXW}^{\otimes n} \Big\Arrowvert P_{U^{n}_{(i)} X^{n}_{(i)} W_{(i)}^n }\right) =2n \delta_n.
\end{equation}
In fact, we have
{\allowdisplaybreaks
  \begin{align*}
 & \mathbb D(   \bar P_{UXW}^{\otimes n} \Arrowvert  P_{U^{n}_{(i)} X_{(i)}^n  W_{(i)}^n }) \\
 & = \mathbb{D}(  \bar P_{X^{n}| U^{n}} \Arrowvert  P_{X_{(i)}^{n}| U_{(i)}^{n}} |\bar P_{U^{n}})\\
&+  \mathbb{D}(\bar P_{W^{n} |X^{n} U^{n}} \Arrowvert P_{W_{(i)}^{n}| X_{(i)}^{n} U_{(i)}^{n}}| \bar P_{X^{n} U^{n}} )
 \end{align*}}
We call $D_1$ and $D_2$ the first and the second term. Then: 
{\allowdisplaybreaks
\begin{align*}
D_1 & {\overset{{(a)}}{=}} \mathbb{D}(  \bar P_{X^{n}}\Arrowvert P_{X_{(i)}^{n}} ) 
{\overset{{(b)}}{=}} \mathbb{D}(  \bar P_{S^{n}}  \Arrowvert P_{S_{(i)}^{n}} ) \stepcounter{equation}\tag{\theequation}\label{divA}\\
& {\overset{{(c)}}{=}} \sum_{j=1}^n \mathbb{D}( \bar P_{ S_{j}|  S^{j-1}}    \Arrowvert  P_{S_{(i),j}| S_{(i)}^{j-1}} | \bar P_{  S^{j-1}} )\\
& {\overset{{(d)}}{=}} \sum_{j \in A_1 \cup A_2} \mathbb{D}(  \bar P_{ S_{j}|  S^{j-1}}    \Arrowvert P_{S_{(i),j}| S_{(i)}^{j-1}}  | \bar P_{S^{j-1}})\\
& {\overset{{(e)}}{=}} \sum_{j \in A_1 \cup A_2} (1- H(S_{j}| S^{j-1}))  {\overset{{(f)}}{\leq}} n \lvert \mathcal V_X \rvert \leq n \delta_n
\end{align*}}
where $(a)$ follows from the fact that $X$ is independent of $U$, $(b)$ from the invertibility of $G_n$,  $(c)$ from the chain rule, 
$(d)$ from \eqref{eq: p11}, $(e)$ from the fact that the conditional distribution $P_{S_{(i),j}| S_{(i)}^{j-1}}$ 
is uniform for $j \in A_1 \cup A_2$ and $(f)$ from Definition \eqref{eq: hz}.

Similarly,
{\allowdisplaybreaks
\begin{align*}
 D_2 & {\overset{{(a)}}{=}} \mathbb{D}( \bar P_{Z^n|X^n U^n}  \Arrowvert P_{Z_{(i)}^n | X^{n}_{(i)} U^{n}_{(i)}}  | \bar P_{X^n U^n })\\
   &  {\overset{{(b)}}{=}} \!\! \sum_{j =1}^n \mathbb D ( \bar P_{Z_j|Z^{j-1} X^n U^{n} } \Arrowvert P_{Z_{(i),j}|Z_{(i)}^{j-1} Z_{(i)}^{n} U_{(i)}^{n}}  | \bar P_{Z^{j-1} X^n U^{n}}  )\\
   &  {\overset{{(c)}}{=}} \!\! \sum_{j \in B_1} \!\!\mathbb D ( \bar P_{Z_j|Z^{j-1} X^n U^{n} } \Arrowvert P_{Z_{(i),j}|Z_{(i)}^{j-1} X_{(i)}^{n} U_{(i)}^{n}}  | \bar P_{Z^{j-1} X^n U^{n}}  )\\
   &  {\overset{{(d)}}{=}} \!\! \sum_{j \in B_1 } ( 1- H(Z_j \mid Z^{j-1} X^n U^{n}) ) {\overset{{(e)}}{\leq}} \delta_n \lvert \mathcal V _{W\mid XU}\rvert 
   \leq n \delta_n,
  \end{align*}}where $(a)$ comes from the invertibility of $G_n$, $(b)$ follows from the chain rule, $(c)$ comes from \eqref{eq: p21}, 
$(d)$ comes from the fact that the conditional distribution $P_{Z_{(i),j}| Z_{(i)}^{j-1} X_{(i)}^{n} U_{(i)}^{n} } $ 
is uniform for $j \in B_1$ and $(e)$ from \eqref{eq: hv}.  Then $D_1+D_2 < 2 n \delta_n$.

Therefore, applying Pinsker's inequality to \eqref{emp1} we have
\begin{equation}\label{emp2}
 \mathbb V(P_{U^{n}_{(i)} X_{(i)}^n W_{(i)}^n }, \bar P_{UXW}^{\otimes n})\leq  2 \sqrt{ \log 2} \sqrt{n \delta_n}:= \delta_n^{(1)} \to 0.
\end{equation}
Note that $Y_{(i)}^n$ is generated symbol by symbol via the channel $\bar P_{Y|X}$.
By Lemma \ref{cuff17}, for each $i \in \llbracket 1,k\rrbracket$,
\begin{equation}\label{oneblockconv}
 \mathbb V (\tilde P_{U^{n}_{(i)} W_{(i)}^n X_{(i)}^n Y_{(i)}^n}, \bar P_{UWXY}^{\otimes n})\!=\!  \mathbb V(\tilde P_{U^{n}_{(i)} W_{(i)}^n }, \bar P_{UW}^{\otimes n}) \! \leq \! \delta_n^{(1)}
\end{equation}
and therefore the left-hand side of \eqref{oneblockconv} vanishes.

Observe that  $V_{(i)}^n$ is generated using $\hat W_{(i)}^n$ (i.e. the estimate of $W_{(i)}^n$ at the decoder)
and not $W_{(i)}^n$.
By the triangle inequality for all $i \in \llbracket 1,k\rrbracket$ 
\begin{align*}
&\mathbb V ( P_{U^{n}_{(i)} \hat W_{(i)}^n X_{(i)}^n Y_{(i)}^n},  \bar P_{UWXY}^{\otimes n} ) \\
 &\leq \! \mathbb V ( P_{U^{n}_{(i)} \hat W_{(i)}^n X_{(i)}^n Y_{(i)}^n},   P_{U^{n}_{(i)} W_{(i)}^n X_{(i)}^n Y_{(i)}^n} ) \stepcounter{equation}\tag{\theequation}\label{triangle}\\
 &+ \! \mathbb V ( P_{U^{n}_{(i)} W_{(i)}^n X_{(i)}^n Y_{(i)}^n}, \bar P_{UWXY}^{\otimes n} ).
\end{align*}We have proved in  \eqref{oneblockconv}  that the second term of the right-hand side in \eqref{triangle} goes to zero, we show that the first term tends to zero as well. 
To do so,  we apply \cite[I.2.6]{Lindvall1992coupling} to 
$$\begin{matrix}[ll]
  A = U^{n}_{(i)} \hat W_{(i)}^n  X_{(i)}^n Y_{(i)}^n & A'=U^{n}_{(i)} W_{(i)}^n X_{(i)}^n Y_{(i)}^n\\
  P = P_{U^{n}_{(i)} \hat W_{(i)}^n  X_{(i)}^n  Y_{(i)}^n} & P'=P_{U^{n}_{(i)} W_{(i)}^n X_{(i)}^n Y_{(i)}^n}\\
\end{matrix}$$
on  $\mathcal A= \mathcal U \times \mathcal W \times \mathcal X \times \mathcal Y$.
Since it has been proven in \cite{arikan2010source} that
\begin{equation*}
p_e:= \mathbb P \left\{\hat W_{(i)}^n  \neq W_{(i)}^n \right\}=O(\delta_n)
\end{equation*}we find that 
$\mathbb V \left( P_{U^{n}_{(i)} \hat W_{(i)}^n X_{(i)}^n Y_{(i)}^n},   P_{U^{n}_{(i)} W_{(i)}^n X_{(i)}^n Y_{(i)}^n} \right) \leq 2 p_e$
and therefore
 {\allowdisplaybreaks
\begin{align*}
&\mathbb V \left( P_{U^{n}_{(i)} \hat W_{(i)}^n X_{(i)}^n Y_{(i)}^n}, \bar P_{UWXY}^{\otimes n} \right)\leq 2p_e+\delta_n^{(1)}=\delta_n^{(2)}\to 0.
\end{align*}}
Since $V^{n}_i$ is generated symbol by symbol from $\hat W^{n}_i$ and $Y^{n}_i$, we apply Lemma \ref{cuff17} again and find 
 {\allowdisplaybreaks
\begin{align*}
&\mathbb V\left(P_{U^{n}_{(i)} \hat W_{(i)}^n X_{(i)}^n Y_{(i)}^n V_{(i)}^n}, \bar P_{UWXYV}^{\otimes n}\right)\leq \delta_n^{(2)}\to 0. \stepcounter{equation}\tag{\theequation}\label{convblock}
\end{align*} }

\subsection{Coordination of all blocks}\label{convallblocks polar}

First, we want to show that two consecutive blocks are almost independent.
To simplify the notation, we set 
 {\allowdisplaybreaks
 \begin{equation*}
\begin{matrix}[ll]
  L:=U^{n} X^n Y^n V^{n}. &
\end{matrix}
\end{equation*}
}

\begin{lem} \label{step2a'}
For $i \in \llbracket 2,k\rrbracket$, we have
$$\mathbb V \left( P_{L_{i-1:i}  \bar C'} , P_{ L_{i-1}\bar C'} P_{L_{i}}\right) \leq  \delta(n).$$
\begin{IEEEproof}
 For $i \in \llbracket 2,k\rrbracket$, we have
{\allowdisplaybreaks
\begin{align*}
 &\mathbb D \left(P_{ L_{i-1:i}\bar C'}  \Arrowvert P_{L_{i-1}  \bar C'}  P_{L_{i}} \right) \\
 &= I (L_{i-1}   \bar C';L_{i})  = I (L_{i};  \bar C' ) +  I (L_{i-1};L_{i}|  \bar C') \\ 
 & {\overset{{(a)}}{=}}  I (L_{i}; \bar C')  
 = I (L_{i};  Z_{(i)} [B'_1] ) 
 {\overset{{(b)}}{=}}  \lvert B' \rvert - H ( Z_{(i)} [B'_1]| L_{i})\\
& {\overset{{(c)}}{=}} \lvert B'_1 \rvert - H (Z[B'_1] |L) +  \delta_n^{(3)}\\
&{\overset{{(d)}}{\leq}} \lvert B'_1 \rvert \!-\!\! \!\sum_{j \in B'_1}\! H (Z_j | Z^{j-1} L)\!+\! \delta_n^{(3)} \\
&{\overset{{(e)}}{\leq}} \lvert B'_1 \rvert - \lvert B'_1 \rvert (1 - \delta_n) +  \delta_n^{(3)} 
 \leq n\delta_n+ \delta_n^{(3)}.
\end{align*}
}To prove $(a)$, observe that, because of the one-time pads on $A_3$ and $B_3$,
$(U^{n}_{(i-1)},  X^{n}_{(i-1)}, Y^{n}_{(i-1)}, V_{(i-1)}^n)$ and 
$(U^{n}_{(i)}, X^{n}_{(i)}, Y^{n}_{(i)}, V_{(i)}^n)$ are dependent only through the recycled common randomness $\bar C'$. 
Therefore, the Markov chain $L_{i-1} - \bar C' - L_{i}$ holds. 
Then,  $(b)$ comes from from the fact that the bits in $B'_1$ are uniform.
To prove $(c)$, note that
{\allowdisplaybreaks
\begin{align*}
 &H (Z_{(i)} [B'_1]| L_{i}) - H (Z[B'_1] |L)\\
 &= H (Z_{(i)} [B'_1] L_{i}) - H (Z[B'_1] L) - H (L_{i}) + H (L)\\
&{\overset{{(f)}}{\leq}}  \delta_n^{(2)} \log{\frac{\lvert \mathcal U \times \mathcal X \times  \mathcal W \times  \mathcal Y \times \mathcal V \rvert}{\delta_n^{(2)}}} \\
& \phantom{\leq} + 
\delta_n^{(2)} \log{\frac{\lvert \mathcal U \times  \mathcal X \times  \mathcal Y \times \mathcal V \rvert}{\delta_n^{(2)}}}\\
&\leq 2 \delta_n^{(2)} (\log{\lvert \mathcal U \times \mathcal X \times  \mathcal W \times  \mathcal Y \times \mathcal V \rvert}-\log{\delta_n^{(2)}}):= \delta_n^{(3)}
\end{align*}}where $(f)$ comes from Lemma \ref{csi2.7} since by \eqref{convblock} we have
\begin{equation*}
 \mathbb V \left( P_{L_i }, \bar P_{UXYV}^{\otimes n}\right) \leq \mathbb V \left( P_{L_i W_{(i)}^n},\bar P_{UWXYV}^{\otimes n}\right) \leq \delta_n^{(2)}
\end{equation*}that  vanishes as $n$ goes to infinity.

Finally, $(d)$ is true because conditioning does not increase entropy and 
$(e)$ comes by definition of the set $B'_1$.
Then we conclude with Pinsker's inequality.
\end{IEEEproof}
\end{lem}

Now that we have the asymptotical independence of two consecutive blocks, to conclude the proof we need the following results.

\begin{lem}\label{step2b'}
We have
$$ \mathbb V \left( P_{ L_{1:k}}, \prod_{i=1}^k  P_{ L_{i}} \right) \leq  \delta(n).$$
\end{lem}

\begin{lem}\label{step2c'}
We have
$$ \mathbb V \left( P_{L_{1:k}}, \bar P_{UXYV}^{\otimes nk}\right) \leq \delta(n).$$
\end{lem} 

We omit the proofs because they are very similar to the proofs of \cite[Lemma 15]{Cervia2018journal} and \cite[Lemma 16]{Cervia2018journal} respectively.

\begin{footnotesize}
\bibliographystyle{IEEEtran}
\bibliography{mybib}

\begin{thebibliography}{10}
\providecommand{\url}[1]{#1}
\csname url@samestyle\endcsname
\providecommand{\newblock}{\relax}
\providecommand{\bibinfo}[2]{#2}
\providecommand{\BIBentrySTDinterwordspacing}{\spaceskip=0pt\relax}
\providecommand{\BIBentryALTinterwordstretchfactor}{4}
\providecommand{\BIBentryALTinterwordspacing}{\spaceskip=\fontdimen2\font plus
\BIBentryALTinterwordstretchfactor\fontdimen3\font minus
  \fontdimen4\font\relax}
\providecommand{\BIBforeignlanguage}[2]{{%
\expandafter\ifx\csname l@#1\endcsname\relax
\typeout{** WARNING: IEEEtran.bst: No hyphenation pattern has been}%
\typeout{** loaded for the language `#1'. Using the pattern for}%
\typeout{** the default language instead.}%
\else
\language=\csname l@#1\endcsname
\fi
#2}}
\providecommand{\BIBdecl}{\relax}
\BIBdecl

\bibitem{cuff2010}
P.~W. Cuff, H.~H. Permuter, and T.~M. Cover, ``{C}oordination capacity,''
  \emph{{IEEE} {T}ransactions on {I}nformation {T}heory}, vol.~56, no.~9, pp.
  4181--4206, 2010.

\bibitem{Cervia2017}
G.~Cervia, L.~Luzzi, M.~Le~Treust, and M.~R. Bloch, ``Strong coordination of
  signals and actions over noisy channels,'' in \emph{Proc. of IEEE
  International Symposium on Information Theory (ISIT)}, 2017.

\bibitem{Cervia2018journal}
\BIBentryALTinterwordspacing
G.~Cervia, L.~Luzzi, M.~Le~Treust, and M.~R. Bloch, ``Strong coordination of
  signals and actions over noisy channels with two-sided state information,''
  2018. [Online]. Available: \url{http://arxiv.org/abs/1801.10543}
\BIBentrySTDinterwordspacing

\bibitem{cuff2011hybrid}
P.~Cuff and C.~Schieler, ``Hybrid codes needed for coordination over the
  point-to-point channel,'' in \emph{Proc. of Allerton Conference on
  Communication, Control and Computing}, 2011, pp. 235--239.

\bibitem{Cervia2017gretsi}
G.~Cervia, L.~Luzzi, M.~Le~Treust, and M.~R. Bloch, ``Polar codes for empirical
  coordination over noisy channels with strictly causal encoding,'' in
  \emph{Colloque GRETSI}, 2017.

\bibitem{cuff2009thesis}
P.~Cuff, ``Communication in networks for coordinating behavior,'' Ph.D.
  dissertation, Stanford University, 2009.

\bibitem{csiszar2011information}
I.~Csisz{\'a}r and J.~K{\"o}rner, \emph{Information theory: coding theorems for
  discrete memoryless systems}.\hskip 1em plus 0.5em minus 0.4em\relax
  Cambridge University Press, 2011.

\bibitem{yassaee2014achievability}
M.~H. Yassaee, M.~R. Aref, and A.~Gohari, ``Achievability proof via output
  statistics of random binning,'' \emph{IEEE Transactions on Information
  Theory}, vol.~60, no.~11, pp. 6760--6786, 2014.

\bibitem{cuff2013distributed}
P.~Cuff, ``Distributed channel synthesis,'' \emph{IEEE Transactions on
  Information Theory}, vol.~59, no.~11, pp. 7071--7096, 2013.

\bibitem{haddadpour2017simulation}
F.~Haddadpour, M.~H. Yassaee, S.~Beigi, A.~Gohari, and M.~R. Aref,
  ``{S}imulation of a channel with another channel,'' \emph{{IEEE}
  {T}ransactions on {I}nformation {T}heory}, vol.~63, no.~5, pp. 2659--2677,
  2017.

\bibitem{elgamal2011nit}
A.~El~Gamal and Y.~H. Kim, \emph{Network information theory}.\hskip 1em plus
  0.5em minus 0.4em\relax Cambridge University Press, 2011.

\bibitem{pierrot2013joint}
A.~J. Pierrot and M.~R. Bloch, ``Joint channel intrinsic randomness and channel
  resolvability,'' in \emph{Proc. of IEEE Information Theory Workshop (ITW)},
  2013, pp. 1--5.

\bibitem{bloch2011physical}
M.~R. Bloch and J.~Barros, \emph{Physical-layer security: from information
  theory to security engineering}.\hskip 1em plus 0.5em minus 0.4em\relax
  Cambridge University Press, 2011.

\bibitem{Lindvall1992coupling}
T.~Lindvall, \emph{Lectures on the Coupling Method}.\hskip 1em plus 0.5em minus
  0.4em\relax John Wiley \& Sons, Inc., 1992. Reprint: Dover paperback edition,
  2002.

\bibitem{treusttech}
\BIBentryALTinterwordspacing
M.~Le~Treust, ``Coding theorems for empirical coordination,'' Tech. Rep., 2015.
  [Online]. Available: \url{https://cloud.ensea.fr/index.php/s/X9e5x8EzJfI7I4Q}
\BIBentrySTDinterwordspacing

\bibitem{csacsouglu2012polar}
E.~{\c{S}}a{\c{s}}o{\u{g}}lu, ``Polar codes for discrete alphabets,'' in
  \emph{Proc. of IEEE International Symposium on Information Theory (ISIT)},
  2012, pp. 2137--2141.

\bibitem{arikan2010source}
E.~Ar{\i}kan, ``Source polarization,'' in \emph{Proc. of IEEE International
  Symposium on Information Theory (ISIT)}, 2010, pp. 899--903.

\bibitem{chou2015secretkey}
R.~A. Chou, M.~R. Bloch, and E.~Abbe, ``Polar coding for secret-key
  generation,'' \emph{IEEE Transactions on Information Theory}, vol.~61,
  no.~11, pp. 6213--6237, 2015.

\bibitem{chou2015polar}
R.~A. Chou and M.~R. Bloch, ``Polar coding for the broadcast channel with
  confidential messages: A random binning analogy,'' \emph{IEEE Transactions on
  Information Theory}, vol.~62, no.~5, pp. 2410--2429, 2016.

\end{thebibliography}
\end{footnotesize}
\end{document}